\RequirePackage{lineno} 
\documentclass[useAMS,usenatbib]{mn2e}

\usepackage{txfonts}
\usepackage[dvips]{graphics}
\usepackage[dvips]{graphicx}
\usepackage{graphicx}
\usepackage{longtable}
\usepackage{color}
\usepackage{url}
\usepackage{dcolumn} 

\newcommand{\OneES}{1ES\,1312$-$423}
\newcommand{\cena}{Centaurus A}

\newcommand{\dgr}{\ensuremath{^\circ}}
\newcommand{\g}{\ensuremath{\gamma}}%
\newcommand{\BeppoSAX}{\textsl{BeppoSAX}}
\newcommand{\ROSAT}{\textsl{ROSAT}}
\newcommand{\fermi}{\textsl{Fermi}-LAT}
\newcommand{\fermiAlone}{\textsl{Fermi}}
\newcommand{\hess}{\textsc{H.E.S.S.}}

\newcommand{\swift}{\textsl{Swift}}
\newcommand{\galex}{\textsl{GALEX}}
\newcommand{\swiftxrt}{\textsl{Swift}/XRT}
\newcommand{\swiftuvot}{\textsl{Swift}/UVOT}
\newcommand{\sync}{synchrotron}

\newcommand{\hessflux}{\ensuremath{\rm \phi(1 TeV)~=~(1.89\pm0.58_{\rm{stat}}\pm0.39_{\rm{sys}})\times 10^{-13}~\rm{cm}^{-2}~\rm{s}^{-1}~\rm{TeV}^{-1}}}
\newcommand{\indexHESS}{\ensuremath{\rm \Gamma~=~2.85\pm0.47_{\rm{stat}}\pm0.20_{\rm{sys}}}}
\newcommand{\hessfluxEdec}{\ensuremath{\phi(E_{\rm 0})~=~(1.18\pm0.35_{\rm{stat}}\pm0.24_{\rm{sys}})\times 10^{-13}~\rm{cm}^{-2}~\rm{s}^{-1}~\rm{TeV}^{-1}}}
\newcommand{\edec}{\ensuremath{E_{\rm 0}~=~1.18~{\rm TeV}}}

\title[\hess\ and \fermi\ discovery of $\gamma$ rays from \OneES]{\hess\ and \fermi\ discovery of $\gamma$ rays from the blazar \OneES}

\author[A.~Abramowski et al.]{
\parbox{\textwidth}{\footnotesize H.E.S.S. Collaboration,
 A.~Abramowski,$^{1}$
 F.~Acero,$^{2}$
 F.~Aharonian,$^{3,4,5}$
 A.G.~Akhperjanian,$^{6,5}$
 E.~Ang\"uner,$^{7}$
 G.~Anton,$^{8}$
 S.~Balenderan,$^{9}$
 A.~Balzer,$^{10,11}$
 A.~Barnacka,$^{12}$
% ...
}\vspace{0.4cm}\\
\parbox{\textwidth}{\tiny
$^1$ Universit\"at Hamburg, Institut f\"ur Experimentalphysik, Luruper Chaussee 149, D 22761 Hamburg, Germany \\
$^2$ Laboratoire Univers et Particules de Montpellier, Universit\'e Montpellier 2, CNRS/IN2P3,  CC 72, Place Eug\`ene Bataillon, F-34095 Montpellier Cedex 5, France  \\
$^3$ Max-Planck-Institut f\"ur Kernphysik, P.O. Box 103980, D 69029 Heidelberg, Germany  \\
$^4$ Dublin Institute for Advanced Studies, 31 Fitzwilliam Place, Dublin 2, Ireland  \\
$^5$ National Academy of Sciences of the Republic of Armenia, Yerevan   \\
$^6$ Yerevan Physics Institute, 2 Alikhanian Brothers St., 375036 Yerevan, Armenia  \\
$^7$ Institut f\"ur Physik, Humboldt-Universit\"at zu Berlin, Newtonstr. 15, D 12489 Berlin, Germany  \\
$^8$ Universit\"at Erlangen-N\"urnberg, Physikalisches Institut, Erwin-Rommel-Str. 1, D 91058 Erlangen, Germany  \\
$^9$ University of Durham, Department of Physics, South Road, Durham DH1 3LE, U.K.  \\
$^{10}$ DESY, D-15735 Zeuthen, Germany  \\
$^{11}$ Institut f\"ur Physik und Astronomie, Universit\"at Potsdam,  Karl-Liebknecht-Strasse 24/25, D 14476 Potsdam, Germany  \\
$^{12}$ Nicolaus Copernicus Astronomical Center, ul. Bartycka 18, 00-716 Warsaw, Poland  \\
% ...
}}

\date{Accepted 2013 June 13}

\pagerange{\pageref{firstpage}--\pageref{lastpage}} \pubyear{2013}

\begin{document}

\label{firstpage}

\maketitle

\begin{abstract}
A deep observation campaign carried out by the High Energy Stereoscopic System (H.E.S.S.) on \cena\ enabled the discovery of $\gamma$ rays from the blazar \OneES, $2^\circ$ away from the radio galaxy. With a differential flux at 1~TeV of $\phi(\rm 1\ TeV)=(1.9\pm0.6_{\rm stat}\pm0.4_{\rm sys})\times 10^{-13}\ \rm{cm}^{-2}\ \rm{s}^{-1}\ \rm{TeV}^{-1}$ corresponding to $0.5 \%$ of the Crab nebula differential flux and a spectral index $\Gamma=2.9\pm0.5_{\rm{stat}}\pm0.2_{\rm{sys}}$, \OneES\ is one of the faintest sources ever detected in the very high energy ($E>100\ {\rm GeV}$) extragalactic sky. A careful analysis using three and a half years of \fermi\ data allows the discovery at high energies ($\rm{E}>100\ \rm{MeV}$) of a hard spectrum ($\Gamma = 1.4 \pm 0.4_{\rm stat} \pm  0.2_{\rm sys}$) source coincident with \OneES. Radio, optical, UV and X-ray observations complete the spectral energy distribution of this blazar, now covering 16 decades in energy. The emission is successfully fitted with a synchrotron self Compton model for the non-thermal component, combined with a black-body spectrum for the optical emission from the host galaxy. 
\end{abstract}


\section{Introduction} \label{intro}

\begin{figure*}
  \centering
	\includegraphics[width=0.45\linewidth]{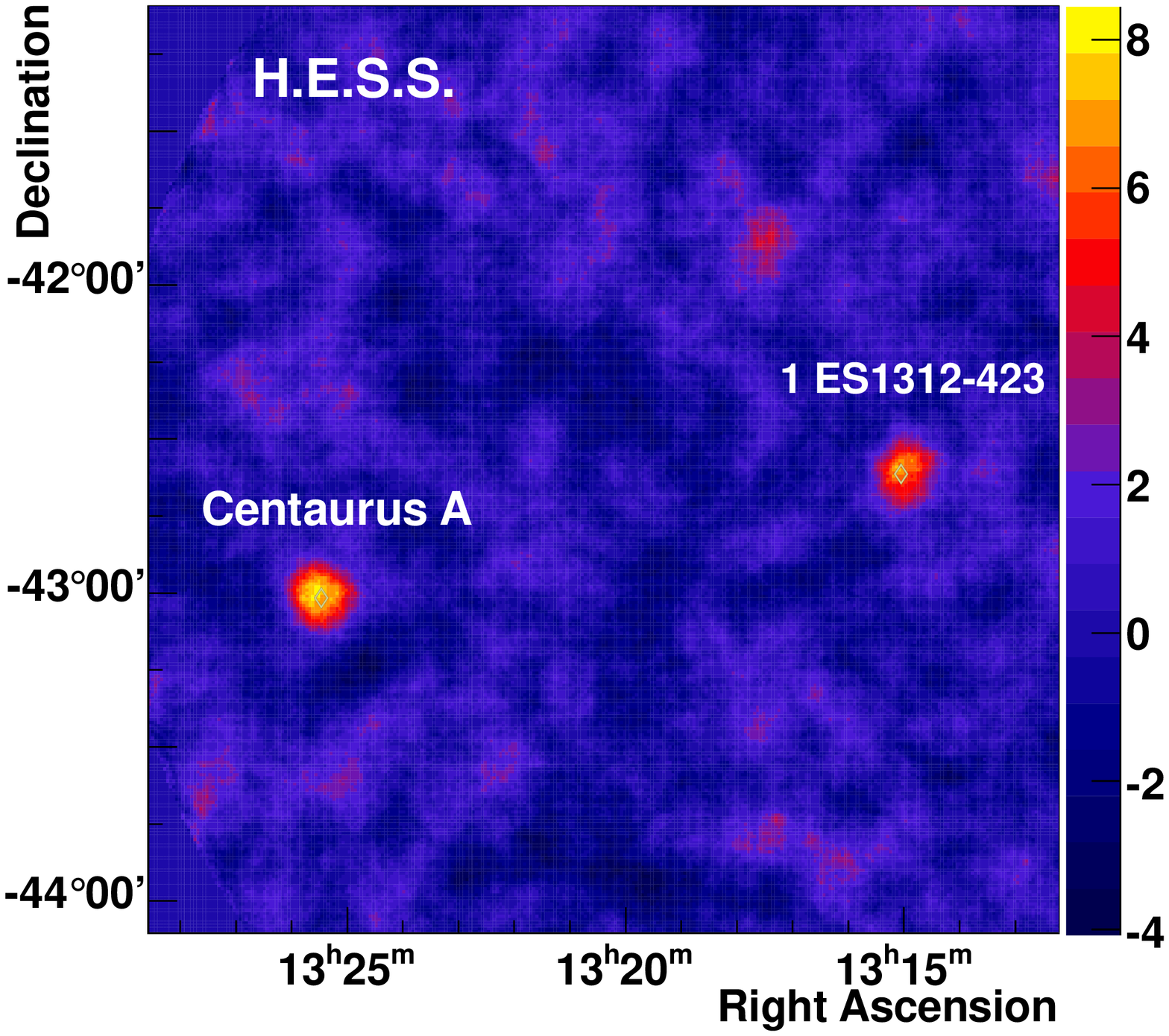} \hfill \includegraphics[width=0.48\linewidth]{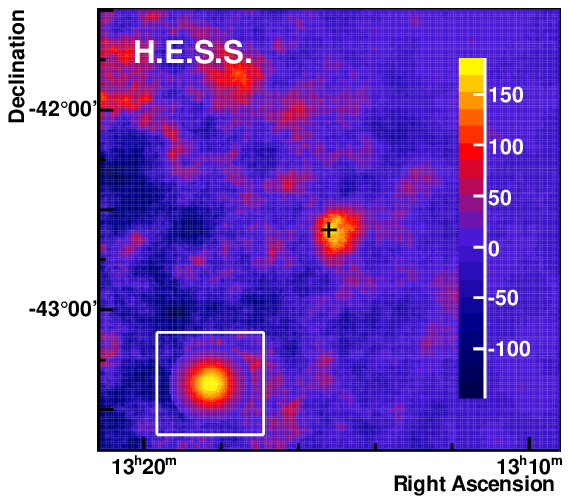}
  \caption{{\it Left}: Significance map of the VHE \g-ray emission of the radio galaxy \cena\ and of the BL Lac \OneES\ in right ascension and declination (J2000). 
    {\it Right}: Map of the excess \g\ rays measured with \hess, smoothed with the point spread function ($68\%$ containment radius of $0.10^\circ$ for these analysis cuts). 
     The cross represents the test position of the source.
     The bottom left inset shows the expected \g-ray excess distribution from a point-like source.}
  \label{SignificanceHESS}
\end{figure*}

BL Lac objects and flat spectrum radio quasars (FSRQs) are the two flavors of active galactic nuclei (AGN) which compose the blazar class. Characterized by a powerful jet aligned at small angles to the line of sight, these two types of blazars have distinct signatures in the optical band: FSRQs exhibit broad emission lines while BL Lac objects show featureless spectra \citep[e.g.,][]{1991ApJ...374..431S,1991ApJS...76..813S,1999ApJ...525..127L}.
The typical spectral energy distribution (SED) of BL Lac objects exhibits two bumps, one at low energy, from radio to X-rays, and the other at higher energies, in the $\gamma$-ray energy band. If the first bump peaks below the infrared to UV domain, a BL Lac can be labeled as a low frequency peaked object (LBL) while it is usually classified as a high frequency peaked object (HBL) if the emission is peaked in the UV/X-ray band \citep{1995ApJ...444..567P}.
Blazar emission models account for the low energy component with \sync\ emission of relativistic electrons accelerated in the jet, while the origin of the high energy bump remains under debate. Leptonic scenarios attribute it to inverse Compton scattering of the electrons off the self-generated \sync\ photon field \citep[synchrotron self Compton or SSC models, e.g.,][]{Band85}, or off externally provided photons e.g.\ from broad-line regions, a dusty torus, or the accretion disk \citep[external Compton or EC models, e.g.,][]{Dermer1993}. The VHE $\gamma$-ray emission in hadronic scenarios can be explained by the interactions of relativistic protons with ambient photons \citep[as, e.g., in][]{1993A&A...269...67M} or magnetic fields \citep[as, e.g., in][]{2000NewA....5..377A}.

Initially detected in X-rays by the Einstein observatory \citep{1990ApJS...72..567G}, 
\OneES\ was subsequently
extracted from optical sky surveys by \citet{1991ApJS...76..813S}, 
who classified it as a BL Lac object. 
Optical imaging by \citet{2000A&A...357...91F} resolved a nucleus ten times fainter than the host galaxy in the R band.
The soft X-ray spectrum derived from \BeppoSAX\ observations by \citet{1998A&A...335..899W} refined the classification of the AGN as a HBL. 
Based on its high ratio of X-ray to radio flux and through a simple modeling, \citet{1996ApJ...473L..75S} 
proposed this low redshift HBL \citep[$z = 0.105\pm0.001$,][]{Rector00} as a potential very high energy (VHE, $E \gtrsim 100$~GeV) \g-ray emitter, 
though with a very faint predicted flux above 1~TeV of 0.7\% of the Crab nebula flux\footnote{The Crab units used in this paper refer to the index $\Gamma$ and differential flux $\phi_0$ at 1~TeV derived by \citet{aha2006} from Crab nebula observations, i.e., $\Gamma = 2.63$ and $\phi_0 = 3.45\times10^{-11}\ \rm{cm}^{-2}\ \rm{s}^{-1}\ \rm{TeV}^{-1}$.}. 

This faint HBL is located at the coordinates $\rm (\alpha_{\rm J2000},\delta_{\rm J2000}) = (13^{\rm h}15^{\rm m}03.4^{\rm s},-42^{\circ}36'50'')$ \citep{2011NewA...16..503M} and it lies at the  edge of the field of view (FoV) of the telescopes of the High Energy Stereoscopic System (\hess) for observations targeted 2\dgr\ away, on the radio-galaxy \cena. In the VHE domain, \cena\ is a faint source (0.8\% of the Crab nebula flux) that was discovered after an extensive \hess\ observation campaign \citep{HESS_CenA}, also unveiling an excess coincident with the position of \OneES\ (see Sec.~\ref{hess})

The \fermiAlone\ Large Area Telescope (\fermi) first and second catalogs, i.e. the 1LAC \citep{2010ApJ...715..429A}, 1FGL \citep{1FGL}, 2LAC \citep{2011ApJ...743..171A}, and 2FGL \citep{2FGL}, do not include a 
counterpart of \OneES\ at high energy (HE, between 100~MeV and 100~GeV). 
However, motivated by the \hess\ detection at VHE, a careful 
modeling of the \cena\ giant lobes emission by the \fermi\ collaboration, using 3.5 years of data, reveals a
faint HE source coincident with \OneES\ (see Sec.~\ref{fermi}).

The HE and VHE spectra derived by the \fermi\ and \hess\ collaborations are combined with multi-wavelength data 
from \textsl{Swift} X-Ray Telescope (XRT) and  Ultra-Violet/Optical Telescope (UVOT), ATOM and ATCA (see Sec.~\ref{swiftXRT}, Sec.~\ref{swift}, Sec.~\ref{atom} and Sec.~\ref{atca}) with the purpose of understanding the properties of the source. 
A standard one-zone SSC model and a black-body emission model for the host galaxy are used to describe the current and the archival data, detailed in Sec.~\ref{mwl}. 
The modeling and its physical implications are discussed in Sec.~\ref{modeling}.

\section{Observations and analysis}\label{ObsMeth}
\subsection{\hess\ dataset and analysis} \label{hess}

\hess\ is an array of four Cherenkov telescopes located 1800 m 
above sea level in the Khomas Highland, Namibia ($23\dgr16'18''$~S, $16\dgr30'01''$~E). 
Each telescope covers a large FoV of 5\dgr\ diameter and consists of a $13$~m diameter optical reflector \citep{Bernlohr} 
and a camera composed of 960 photomultipliers \citep{Vincent}. 
The coincident detection of a Cherenkov flash from an extended air shower with at least two telescopes triggers 
the acquisition of its images and allows a good cosmic-ray background rejection above $\sim$100~GeV \citep{Funk}. 

Located 2\dgr\ away from \cena, \OneES\ benefits from the intensive observation campaign on this FoV. 
The dataset studied in this paper is selected with standard quality criteria \citep [stable detector and good weather, as described in][]{aha2006}, 
which yielded 150.6 hours exposure time from April 2004 to July 2010 at an average zenith angle of $24^{\circ}$. 
The correction of this total exposure time for the decrease in efficiency due to the large offset ($\sim$2\dgr) 
of the source with respect to the camera center leads to a total corrected exposure of 48.4 hours. This decrease in efficiency 
does not bias the analysis of the dataset, as shown in appendix~\ref{Appendix}.

Data are analyzed using the analysis method described in \citet{Becherini} and cross-checked with the method of \citet{Naurois}. These methods both achieve enhanced background rejection and sensitivity at low energies with respect to standard analysis methods \citep[e.g.,][]{aha2006}. The first procedure, used to derive the results shown in this paper, is based on a Boosted Decision Tree technique, with a multivariate combination of discriminant parameters from the Hillas \citep[see, e.g.,][]{aha2006} and the 3D-model \citep{2006APh....25..195L} analysis methods.

Both analysis methods are applied using a minimum image intensity of 60 photoelectrons (p.e.) 
yielding a threshold energy of $280 \;\rm GeV$, and give consistent results.
The VHE significance map of \OneES\ showing the presence of the two AGN in the same FoV (see Fig.~\ref{SignificanceHESS}) 
is obtained with the \textsl{Ring} background modeling method \citep{2007A&A...466.1219B} and with an inner ring radius of $0.7^{\circ}$.
The smoothed, background-subtracted map of the number of \g\ rays observed around the position of \OneES, obtained with the same background modeling method,
is shown in the right-hand panel of Fig.~\ref{SignificanceHESS}.  

\begin{figure}
  \centering
  \includegraphics[width=9cm]{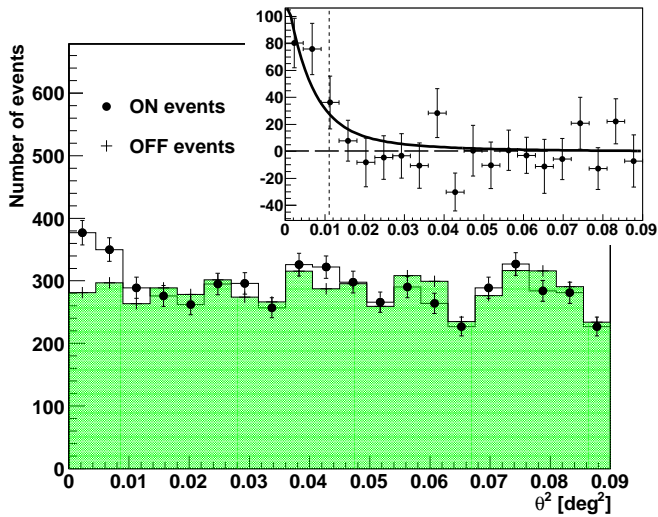}
  \caption{The distribution of the squared angular distance between the test position and the reconstructed shower direction
    for ON-source events (black points) and normalized OFF-source events (green shaded area). 
    In the inset: background-subtracted distribution of the squared angular distance in deg$^2$ between the fitted position and the reconstructed shower direction, representing the excess events from this source. The superimposed line is a fit of the PSF to the data. The PSF 68\% containment radius of 0.102\dgr\ is shown as a vertical dashed line.}
  \label{fig:subtheta2}
\end{figure}

The distribution of the squared angular distance $\theta^2$ 
between the reconstructed shower direction and the test position, obtained by projecting the two-dimensional maps on the radial direction, is shown in Fig.~\ref{fig:subtheta2}, 
for ON-source and normalized OFF-source events.
The distribution of the excess, shown in an inset, is compatible with the \hess\ point spread function 
(PSF, black line in the inset on Fig.~\ref{fig:subtheta2}). The total excess, obtained with the \textsl{Reflected} background modeling method \citep{aha2006} within a radius of 0.102\dgr\ (PSF 68\% containment radius),
corresponds to $149 \pm 28$~events\footnote{The total numbers of ON and OFF-source events are $\rm N_{ON} = 780$
and $\rm N_{OFF} = 30120$, respectively, with a background-normalization 
factor $\alpha \simeq 0.0209$.} at the test position of the source, for an overall significance\footnote{Since \OneES, whose location is precisely measured, is a known candidate \g-ray emitter \citep{1996ApJ...473L..75S}, no trial factor is accounted for in the significance.} of $5.7\sigma$. 

The fit of a point-like source model convolved with the \hess\ PSF to the excess events 
locates the emission at 
$\rm (\alpha_{\rm J2000},\delta_{\rm J2000}) = (13^{h} 14 ^{m} 58.5^{s} \pm 4.2^{s}_{\rm stat} \pm 1.3 ^{s}_{\rm sys},-42^{\circ} 35' 49'' \pm 48''_{\rm stat} \pm 20''_{\rm sys})$, which is compatible with the test position at the $1\; \sigma$ level. 

The differential energy spectrum $\phi(E) = {\rm d}N / {\rm d}E$ of the VHE \g-ray emission is  derived above $\rm 280\;GeV$ with a forward folding technique \citep{2001A&A...374..895P}. 
A fit with a power law $\phi(E) = \phi(E_{\rm 0}) \times (E/E_{\rm 0})^{-\Gamma}$ yields best-fit parameters \indexHESS\ and \hessfluxEdec\
at the decorrelation energy \edec. 
This corresponds to a differential flux at $\rm 1\;TeV$ 
of \hessflux, equivalent to $0.5\%$ of the Crab nebula differential flux. 
\OneES\ is thus one of the faintest extragalactic sources ever detected in the VHE band. 

\begin{figure}
   \hspace{-0.3cm}
   \includegraphics[width=9.4cm]{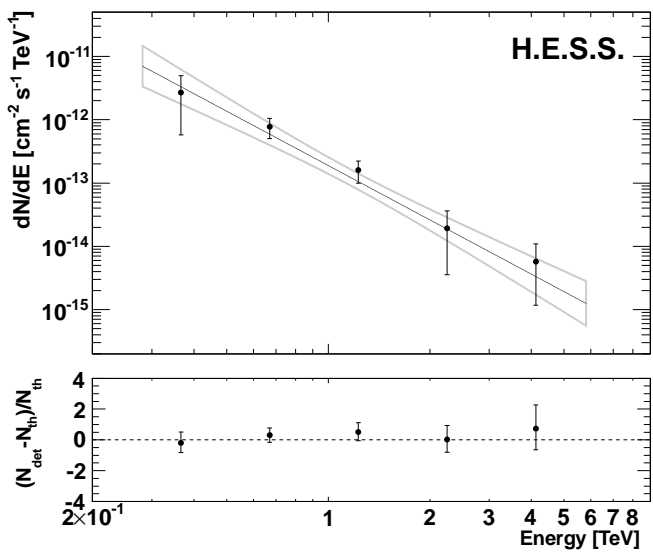}
      \caption{\hess\ differential energy spectrum of \OneES. 
        The gray butterfly represents the $1\sigma$ contour 
        of the best power-law model that fits the data in the observed energy range (280 GeV$-$5.8 TeV). The lower panel shows the fit residuals, i.e. (N$_{\rm det}$ - N$_{\rm th}$)/N$_{\rm th}$, where N$_{\rm det}$ and N$_{\rm th}$ are the detected and expected number of excess events, respectively.}
   \label{fig:spec}
   \end{figure}

The $1\sigma$ confidence contour of the power-law fit, referred to as the ``butterfly'', 
together with the spectral points and residuals, are shown in Fig.~\ref{fig:spec}. 
The numbers of excess events $N_{i,\rm det}$ detected in the five energy bins $i$ do not significantly differ from the numbers of expected events $N_{i,\rm th}$, yielding $X^2 = \sum_i (N_{i,\rm det} - N_{i,\rm th})^2/\sigma_{N_{i,\rm det}}^2$ of 1.03 with 3 degrees of freedom.

\begin{figure*}
	\centering
  \includegraphics[width=0.95\linewidth]{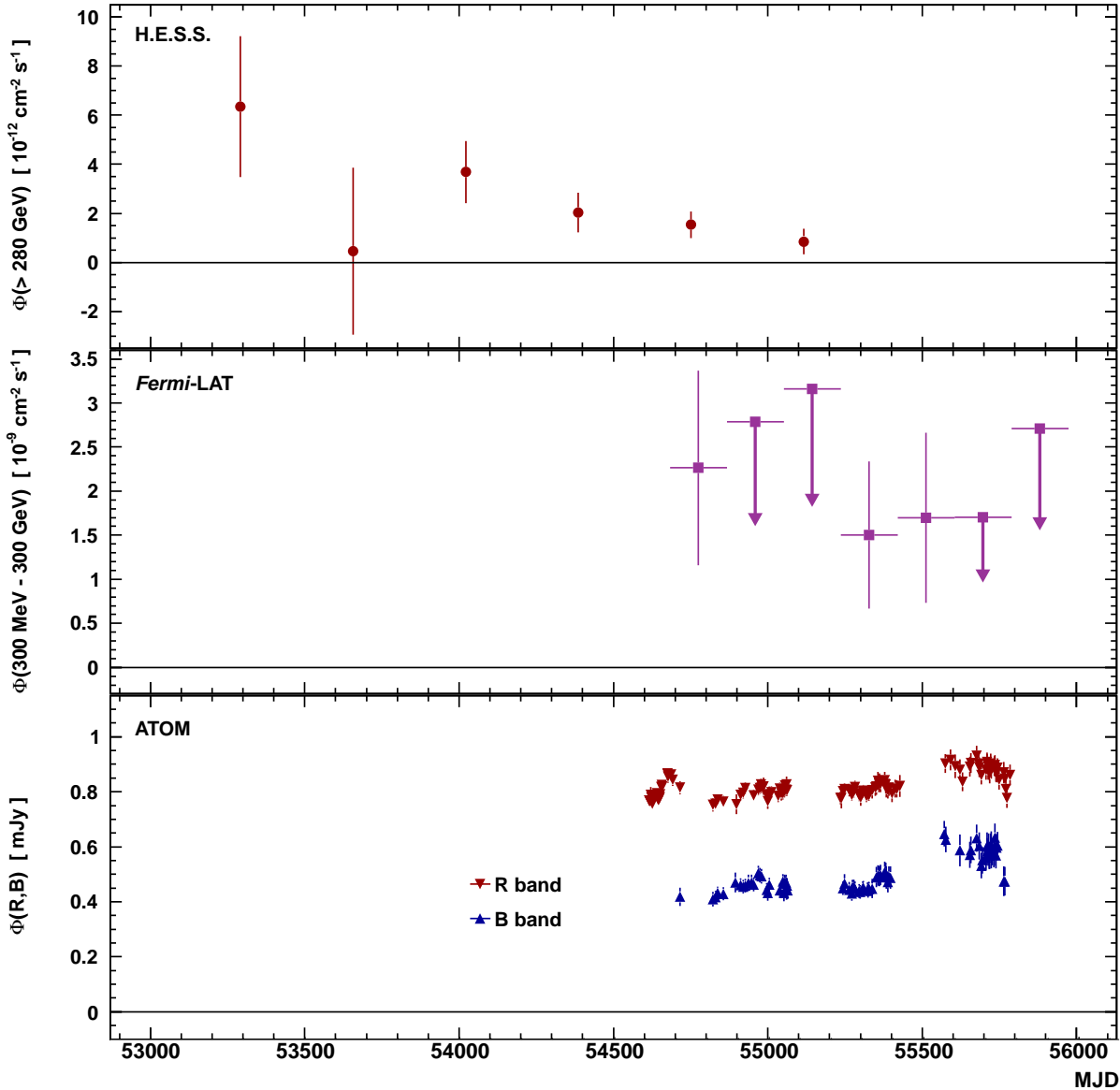}
  \caption{{\it Top:} Integral VHE \g-ray flux of \OneES\ above the threshold energy measured with \hess\ {\it Middle:} Integral HE \g-ray flux of \OneES\ in the \fermi\ energy band. 95\% CL upper limits are computed for the time bins with $\rm{TS} <4$. {\it Bottom:} Dereddened flux measured by ATOM in the R and B bands, represented with downward and upward pointing triangles, respectively.}
\label{fig:lc}
\end{figure*}

The VHE light curve, computed on a year-by-year time scale, is shown in the top panel of Fig.~\ref{fig:lc}. Since the number of excess events is statistically low, a smaller temporal binning does not provide more information on variability. A search for flux variability is carried out using the fractional variance $F_{\rm var}$ \citep[see, e.g.,][]{Vaughan}, an estimator of the intrinsic variance normalized to the square of the mean flux, which is computed by quadratically subtracting the contribution of the experimental uncertainties from the observed variance.

The measured normalized excess variance of $F_{\rm var}^2 = 0.17\pm0.44$ is compatible with zero, indicating that any potential variability is washed out by measurement uncertainties.  A 99\% confidence level (CL) upper limit, calculated using the method of \citet{Feldman} under the hypothesis of a Gaussian uncertainty, is $F_{\rm var}^2 \leq 1.30$, which corresponds to $F_{\rm var}(\mathrm{year}) \lesssim 1.1$. This limit means that the VHE \g-ray flux of \OneES\ did not vary on average by more than a factor 1.1 on the time scale of one year.

\subsection{\fermi\ dataset and analysis} \label{fermi}

The LAT on board the \textsl{Fermi} satellite
is a pair-conversion telescope designed to detect \g\ rays from
$\rm 20 \;MeV$ up to energies greater than $300 \; GeV$. The characteristics and performance of
the instrument are described in \cite{atw09}. The LAT observes the full sky
every 3 hours (two orbits) and each source is in the FoV for $\sim 30$ minutes during this period.

The LAT data on \OneES\ analyzed hereafter span 3.5 years, from August 4, 2008 (MJD 54682) to February 16, 2012 (MJD 55973). 
Only events with a high probability of being photons
(belonging to the SOURCE class), with zenith angles less than
$100^\circ$ and with reconstructed energies between 300 MeV and 300
GeV are retained. The {\tt P7SOURCE$\_$V6} Instrumental Response Functions
(IRFs) are used to describe the detector and the data are analyzed
with the ScienceTools v9r23p1. 

A binned likelihood analysis chain \citep{1996ApJ...461..396M,BSL}, implemented in the {\tt gtlike} tool, is used to best match the spectral model with the
front and back events\footnote{See section 2.2.1 of \citet{atw09} for definition of front and back events.}, which are analyzed separately to maximize the sensitivity. 

The analysis is carried out on a region of interest (ROI) of $15^\circ$ around the \OneES\ coordinates, where the events are grouped in $0.1^\circ\times0.1^\circ$ bins and using ten energy bins per decade between 300~MeV and 300~GeV. The sky model is constructed using the standard model for Galactic interstellar diffuse emission, 
an isotropic background
component \footnote{The backgrounds are described in the files {\tt gal$\_$2yearp7v6$\_$v0.fits} and {\tt
iso$\_$p7v6source.txt} available from the FSSC \url{http://fermi.gsfc.nasa.gov/ssc/}} and the sources of the 2FGL catalog \citep{2FGL}, with the spectral models derived therein. The parameters of the sources close to \OneES\ ($\leq 3^\circ$) are left free during the fitting procedure, while the parameters of more distant sources are frozen to their 2FGL values. 
The extended \g-ray emission from the \cena\ lobes is modeled using a spatial template based on the 22 GHz WMAP image of the region
\citep{hinshaw2009}, as in \citet{Cenalobes}. The validity of the model is checked by subtracting the predicted count map from the observed one, yielding no significant residuals.
A point-like source is added to the sky model at the test position of \OneES, whose spectrum is described with a power law. The positions of the sources are frozen to the input values during the minimization process.

With a Test Statistic of the likelihood analysis of 32.6, approximately 5.7$\sigma$, \OneES\ is detected
by the LAT with a flux of $(1.59\pm0.48_{\rm stat}\ ^{+ 0.14}_{-0.54\ \rm sys} ) \times 10^{-15}$ photons cm$^{-2}$\,s$^{-1}$\,MeV$^{-1}$ at the decorrelation energy $E_{\rm 0}=22.5$~GeV  and a photon index $\Gamma=1.44\pm0.42_{\rm stat}\ ^{+ 0.20}_{-0.25\ \rm sys}$, where the systematic uncertainties are evaluated using the bracketing IRFs method \citep{2012ApJS..203....4A}. The low average flux measured with \fermi\ using 3.5 years of data is consistent with the non-detection ($\rm{TS}<25$) of \OneES\ in the second catalog (interpolated TS of $\sim19$ assuming a steady-state emission). The upper end of the energy range covered by \fermi\ is set to 300~GeV since the three highest photon energies measured in a 95\% containment radius are 102~GeV, 181~GeV and 294~GeV. Four flux points are computed in energy bins of equal width in log scale by performing a {\tt gtlike} analysis with the photon index of the source frozen to its best-fit value. The same analysis is performed to compute the light curve on time intervals of 6 months, shown in the middle panel of Fig.~\ref{fig:lc}. Upper limits at the 95\% CL are computed for spectral and temporal bins with $\rm{TS}<4$. The variability indicator provided by the likelihood ratio method in \cite{2FGL} is computed as ${\rm TS_{var}} = 13.1$ for six degrees of freedom, showing no significant variations with an equivalent chance probability of 41\%.

\subsection{\swiftxrt\ dataset and analysis} \label{swiftXRT}

The \hess\ collaboration triggered an observation of \OneES\ with the space-based \swift\ X-ray observatory
\citep{2005SSRv..120..165B}, performed on January 25, 2011 at 01:22 UTC (ObsID 00031915001) with 4.7
ks exposure time. The photon-counting (PC) mode data are processed with the standard {\tt xrtpipeline} tool
({\small \tt HEASOFT 6.12}), with the source and background-extraction regions defined as a 20-pixel ($\sim 4.7\ {\rm arcsec}$) radius
circle, the latter being centered nearby the former without overlapping. The source region count rate is $\sim
0.7\,{\rm counts}\,{\rm s^{-1}}$, a rate considered to be at the limit of risk of pile up, but a King function
fit to the PSF shows no evidence for pile up in the inner part of the source region. Also, the results of the spectral analysis are compatible within errors to those found with a source region defined
by an annulus of 20-pixel outer radius and 4-pixel ($\sim 0.9\ {\rm arcsec}$) inner radius, thus excluding 58\% of the events. 

The {\tt xrtmkarf} tool is
used to generate a dedicated Ancillary Response Function (ARF) at the location of the source in the FoV, along
with the latest spectral redistribution matrices from CALDB. The \swiftxrt\ spectrum is rebinned to have at least 20 counts per bin using {\tt grppha}, yielding a usable
energy range between $0.3$ and $7.0\,{\rm keV}$. Multiple model spectra are tested with {\small
  \tt PyXspec  v. 1.0.1}, the response functions {\tt swxpc0to12s6\_20010101v013} and the dedicated ARF. Systematic errors on the \swiftxrt\
spectra and absolute flux are less than $3\%$ and $10\%$ respectively \citep{2009A&A...494..775G}.

\begin{table*}
\centering
\begin{tabular}{*{9}{c}}
   \hline
Model & $N_{\rm H}\times 10^{20}$&  $\phi_0$                                       & $\Gamma$ or $a$ or $\Gamma_{0}$ & $b$ or $\Gamma_{1}$  & $E_{\rm break}$ &  $\chi^2/ndf$ & $P_{\chi^2}$ & LRT($H_0 = \rm{PWL}$) \\
      & ${\rm cm}^{-2}$         &  $10^{-3} {\rm cm}^{-2}\ \rm{s}^{-1}\ \rm{keV}^{-1}$ &       &        &  keV    &               &              &               \\ 
   \hline
PWL    & $7.8$     & $5.7\pm0.1$               & $1.91\pm0.02$                          &      --       &  --  &  155.0/122     &   23\%       &     --        \\ 
BPWL   & $7.8$     & $5.9\pm0.1$               &$1.67\pm0.07$                    & $2.18\pm0.08$ & $1.6\pm0.2$ &  124.8/120      &  36\%       &  $5.1\sigma$             \\ 
LP     & $7.8$     & $6.1\pm0.1$               & $1.73^{+0.04}_{-0.05}$                 & $0.47\pm0.09$ &  --  & 122.4/121     &  45\%       &    $5.7\sigma$ \\ 
LP - min $N_{\rm H}$     & $6.7$     & $5.9\pm0.1$   & $1.66\pm0.04$                          & $0.55\pm0.09$ &  --  & 122.4/121     &   45\%       &   --          \\ 
LP - max $N_{\rm H}$    & $9.1$     & $6.3\pm0.1$   & $1.81\pm0.04$                           & $0.38\pm0.09$ &  --  & 122.6/121     &    44\%      &    --         \\ 
LP - free $N_{\rm H}$    & $7.3\pm3.6$ & $6.0\pm0.7$   &  $1.70\pm0.23$      & $0.51\pm0.27$ &  --  & 122.3/120     &   42\%      &    $5.4\sigma$        \\ 
PWL - free $N_{\rm H}$  &   $14.2\pm 1.3$        & $7.3\pm0.4$       & $2.14\pm0.05$     &      --       &  --  &    125.8/121  &  37\% &  $5.4\sigma$ \\
\hline
\end{tabular}
\caption{Fit parameters and uncertainties for the spectral models and column density values studied in Sec.~\ref{swiftXRT}. The parameters described in column 4 and 5 depend on the model (power law, log parabola, borken power law) as discussed in the text. The $\chi^2$, numbers of degrees of freedom, and corresponding probabilities are shown in columns 7 and 8. The last column indicates the significance of the fit improvement for models nested with the PWL model with fixed column density.}
\label{tab:xspecfits}
\end{table*}

The XRT spectrum is first studied using the weighted average column density of Galactic HI $N_{\rm H}=7.8\times10^{20}\,{\rm cm^{-2}}$ that is extracted from the Leiden/Argentine/Bonn (LAB) survey
\citep{Kalberla2005} with the $N_{\rm H}$ tool from HEASARC\footnote{http://heasarc.gsfc.nasa.gov/cgi-bin/Tools/w3nh/w3nh.pl}. 
The best-fit power-law model $\phi(E)=\phi_0 (E/E_0)^{-\Gamma}$, PWL in Table~\ref{tab:xspecfits}, yields a reduced $\chi^2$ of 1.1, with
deviations from zero in the residuals at both low and high energies. 
A likelihood ratio test (LRT) prefers the simplest
smoothly-curved function, a log parabola (LP) $\phi(E)= \phi_0 (E/E_0)^{-a - b\log(E/E_0)}$,
at the $5.7\sigma$ level with a reduced $\chi^2$ of 1.01. 
The best-fit curvature parameter $b$ measured in the spectrum of \OneES\ 
is characteristic of the TeV candidates
observed in X-rays as noted by \citet{2008A&A...478..395M,2011ApJ...739...73M}.  
A broken power law (BPWL), $\phi(E)=\phi_0 (E/E_0)^{-\Gamma_{0}\times\Theta(E<E_{\rm break})-\Gamma_{1}\times\Theta(E>E_{\rm break})}$ 
where $\Theta$ is the Heaviside function, is also an acceptable model with a reduced $\chi^2$ of 1.04, and a break of $\Delta \Gamma=\Gamma_{1}-\Gamma_{0} \simeq 0.51 \pm 0.11$ consistent
with those usually found in X-ray selected BL Lac type objects \citep{1996ApJ...463..444S,1996ApJ...463..424U,1998A&A...335..899W}.
The log parabola has one parameter fewer than the broken power law and is therefore used in the following.

\begin{figure}
     \centering
	   \includegraphics[width=8.4cm]{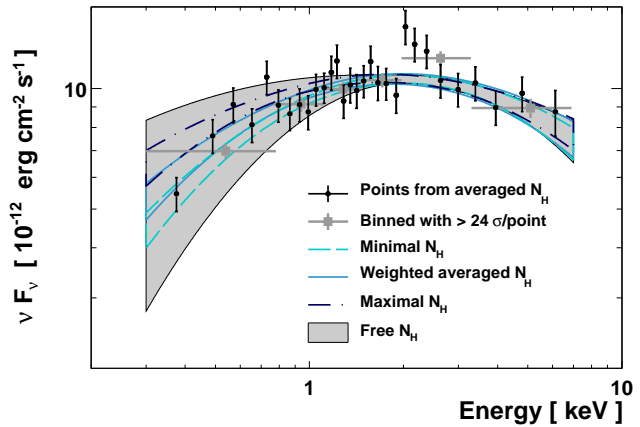}
      \caption{\swiftxrt\ spectrum of \OneES. 
        The column density of hydrogen $N_{\rm H}$ varies greatly over the 1\dgr$\times$1\dgr~FoV, 
        three values of $N_{\rm H}$ (dashed line for minimal, solid line for weighted average and dotted/dashed line for maximal) 
        are used to deabsorb the spectrum. The gray butterfly represents the best log-parabola model 
        that fits the data in the observed energy range (0.3 keV$-$7.0 keV) for a free column density.}
   \label{fig:XRTspec}
   \end{figure}

In addition to the uncertainties on the optical depth correction for the spin temperature of the gas \citep[see, e.g., the discussion in][]{2010arXiv1002.0081J}, the values of $N_{\rm H}$ from the LAB survey show significant variations within one degree of the source, ranging from a minimal value of 
$N_{\rm H}=6.7\times10^{20}\,{\rm cm^{-2}}$ to a maximal value of $N_{\rm H}=9.1\times10^{20}\,{\rm cm^{-2}}$. 
This could indicate that fluctuations of the interstellar medium on scales smaller than the LAB survey's
half-power beam-width (HPBW) of $\simeq 0.6^\circ$ exist but are barely detectable.
Checking for fluctuations on smaller scales in the \textsl{IRAS} $100\,{\mu m}$
\footnote{using the tool http://irsa.ipac.caltech.edu/applications/DUST/} dust maps, which have a better resolution and correlate extremely well
at high Galactic latitudes with maps of HI emission \citep{1998ApJ...500..525S}, no evidence for large
fluctuations in a 5 arcmin radius centered on \OneES\, is found. Following the relation of
\cite{2009MNRAS.400.2050G}, the estimated total extinction in the V band of A(V) = $0.336 \pm 0.016\,{\rm mag}$ corresponds to
$N_{\rm H} = {\rm A(V)}\times(2.21\pm0.09)\times10^{21}\,{\rm \ cm}^{-2}= (7.4\pm0.5)\times10^{20}\,{\rm \ cm}^{-2}$. 
The minimal and maximal $N_{\rm H}$ values of the LAB survey are compatible with $\pm 3\sigma$
deviations from this value, and are hence used as conservative limits on the column density in the following.
This slightly changes the best-fit parameters of the log-parabolic spectrum (fourth and fifth line in Table~\ref{tab:xspecfits}),
the low-energy slope $a$ being most affected by these variations, as shown in Fig.~\ref{fig:XRTspec}.

For reference, leaving the column density as a free parameter with the log-parabolic model
yields a flat X-ray spectrum in $\nu{F}_\nu$ up to $\sim 1\,{\rm keV}$, marginally preferred  to
a power-law model with free column density (a likelihood ratio test yielding a $1.9\sigma$ improvement).
The latter would suggest additional absorbing material with a column density $\geq 7\times 10^{20}\,{\rm\ cm}^{-2}$,
for which little or no evidence has been found so far in BL Lac objects \citep{Perlman}. 
The curvature, which is compatible with values usually found for X-ray bright BL Lac objects, appears to be intrinsic 
to the source and the first bump in the SED of \OneES\ should then peak in the energy band covered by
\swiftxrt.

\subsection{\swiftuvot\ analysis} \label{swift}

Simultaneously with \swiftxrt\ observations, the \swiftuvot\ \citep{RomingUVOT} took six snapshots of
\OneES\ with the filter {\it uvm}2 (224.6~nm). They are integrated with the \texttt{uvotimsum} tool of the
package {\small \tt HEASOFT 6.12} and analyzed using the \texttt{uvotsource} tool, with circular ON and OFF
regions of radius 5 arcsec. The background is estimated from different OFF regions, at least 25 arcsec away
from the source. A change of the background regions impacts the reconstructed flux at the percent level.  The
magnitudes are dereddened according to the extinction laws of \citet{2003ARA&A..41..241D} with a total absorption at
224.6~nm A$_\lambda = $ 1.1 mag corresponding to the \textsl{IRAS} extinction mentioned in Sect.~\ref{swiftXRT}.  The
flux densities are computed from the magnitude according to the zero points of \citet{uvotphot}.  The source
does not show any sign of variability on the hour time scale ($F_{\rm var}^2(\mathrm{hour}) <
0.012$ at the 99\% confidence level), with an average flux at 224.6~nm of $(3.19 \pm 0.08) \times 10^{-12}$ erg cm$^{-2}$
s$^{-1}$. Following \citet{2007A&A...467..501T}, a conservative systematic uncertainty on the UV flux of 15\% is adopted.

A joint fit of the UVOT measurement with the X-ray spectrum, adding the interstellar extinction
directly to the model with the {\tt redden} component set to the \textsl{IRAS} value, does not significantly change the
results from Sect.~\ref{swiftXRT}, nor the estimate of the \swiftuvot\ flux.

\subsection{ATOM dataset and analysis}
\label{atom}

Optical observations were performed with the ATOM telescope \citep[Automatic Telescope for Optical
  Monitoring,][]{Haus04} at the \hess\ site from May 2008 to August 2011.  Absolute flux values are
calculated using differential photometry with 4~arcsec radius aperture for all filter bands, against four stars calibrated with 
photometric standards.

The 83 and 122 measurement points in the B and R bands, respectively, show small flux variations during the two
years of observation, with fractional variances of $10.6\pm0.9\%$ and $4.0\pm0.4\%$, respectively. The
observed average magnitudes $m_B = 17.87\pm0.02_{\rm stat}\pm0.04_{\rm sys}$ and $m_R = 16.66\pm0.01_{\rm
  stat}\pm0.02_{\rm sys}$ are converted to fluxes using absorptions of
A$_\lambda(B) = $ 0.499 mag and A$_\lambda(R) = $ 0.299
mag \citep{2003ARA&A..41..241D} and standard zero points \citep{Bessel90}, yielding a total flux from the host and the
nucleus of $\phi(B) = (3.33\pm0.05_{\rm stat}\pm0.10_{\rm sys}) \times 10^{-12}\ {\rm erg}\ {\rm cm}^{-2}\ {\rm s}^{-1}$,
$\phi(R) =(3.47\pm0.02_{\rm stat}\pm0.06_{\rm sys}) \times 10^{-12}\ {\rm erg}\ {\rm cm}^{-2}\ {\rm s}^{-1}$, the systematic
uncertainty arising from flat fielding as well as dark and bias corrections. The ATOM light-curves in the B
and R bands are shown in the bottom panel of Fig.~\ref{fig:lc}.

\subsection{ATCA observations}
\label{atca}

Radio observations of \OneES\ were made by the Australia Telescope Compact Array \citep[ATCA;][]{2011MNRAS.416..832W} on May 10, 2012, using the array configuration EW352. Flux densities are calibrated against PKS~1934--638, the ATCA primary flux density calibrator. These observations were made with 2~GHz bandwidths provided by the Compact Array Broadband Backend \citep[CABB;][]{2011MNRAS.416..832W}, centered on 5.5, 9.0, 17, 19, 38 and 40~GHz. At the lower four frequency bands, the source has a flux density of 9 mJy. It has a flux density of 6.5~mJy at 38 GHz and 4.6 mJy at 40 GHz. Thus this source has a relatively flat radio spectrum with a spectral index of $0.20\pm0.05$. Uncertainties at the 3$\sigma$ level on the flux density are conservatively estimated at 2~mJy.

The pointing model was updated for the 17/19~GHz and 38/40~GHz band observations by a five point cross scan on a nearby bright AGN. Scans at each frequency were 2 minutes in length and the source was at elevation 48 degrees at 5/9~GHz and a few degrees higher at the highest frequencies, thus minimizing atmospheric effects. Data reduction followed standard procedures as described in 
\citet{2012arXiv1205.2403S}. There are no signs of the source being extended on the $L=4.4~{\rm km}$ maximum baselines in this array, for which the angular resolution goes down to $0.4~{\rm arcsec}$ at 40~GHz.

\section{Modeling of the spectral energy distribution}\label{SED}

\subsection{Comparison of archival and current data}\label{mwl} 

\begin{figure}
  \centering
  \includegraphics[width=8.4cm]{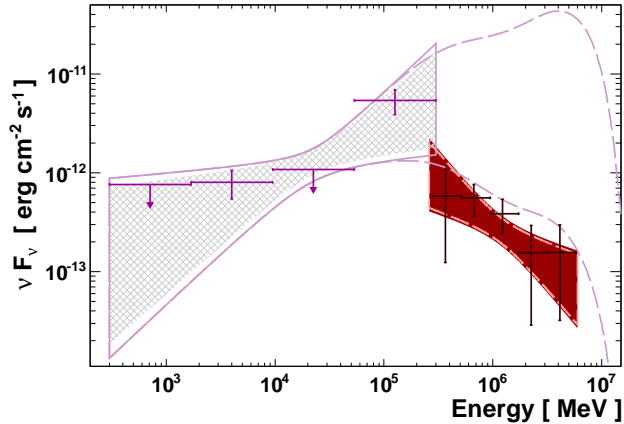}
  \caption{\fermi\ and \hess\ spectra, represented by the empty magenta and filled dark red butterflies respectively. 
    The \hess\ (resp. \fermi) spectrum includes statistical and systematic errors on the flux and index, 
    for reference the 1$\sigma$ statistical butterfly is shown in dashed light red (resp. hatched light magenta). 
    The \fermi\ butterfly is extrapolated to VHE (dashed magenta), 
    taking into account the EBL absorption according to \citet{Fran08}.}
\label{fig:FermiAndHESS}
\end{figure}

The $1\sigma$ \fermi\ butterfly ($300\ \rm{MeV}<E<300\ \rm{GeV}$) and the \hess\ spectrum are represented by the empty and filled dark butterflies in Fig.~\ref{fig:FermiAndHESS}. The \hess\ butterfly includes the systematic uncertainties on the index ($\sim 0.20$) and on the flux ($\sim 20\%$), added  quadratically at the decorrelation energy to the statistical one. The systematic uncertainties on the \fermi\ spectral parameters are propagated in the same way.
The \fermi\ spectral extension to VHE (dashed line), which is absorbed by the extragalactic background light (EBL)
according to \citet{Fran08}, overshoots the H.E.S.S. $1\sigma$ confidence contour. The difference between the spectral indexes is explained by intrinsic curvature, supported by the empirical relation of \citet{SEDfermiBlazars} between the peak frequency $\nu_{\rm peak}^{\rm IC}$ of the high energy bump
and the HE photon index $\Gamma$, which reads $\displaystyle \log \nu_{\rm peak}^{\rm IC}[\rm Hz] = -4\Gamma + 31.6 = 26.2 \pm 1.4$, 
corresponding to a peak energy between $\sim25$~GeV and $\sim15$~TeV. The low statistics of the \fermi\ data at high energies (three photons above 100~GeV) does not enable a straightforward characterization of this peak energy. 
The \fermi\ and \hess\ fluxes differ by a factor $r \sim 6$ at 300~GeV, with $\log_{10} r = 0.8\pm0.6_{\rm stat}\pm0.3_{\rm sys}$, but the large statistical and systematic uncertainties do not suggest a significant discrepancy. Though no flux variations can be detected within the uncertainties, a more significant mismatch could be explained by the non-simultaneous sky coverage of \hess\ and \fermi, as shown in Fig~\ref{fig:lc}. Since little variability in the spectral index is found at HE, irrespective of the blazar class \citep{2010ApJ...710.1271A}, the emission model mostly aims at reproducing the slope observed at HE, and the VHE spectrum is used as a baseline for the normalization. 

\begin{figure*}
	\centering
   \includegraphics[width=18.5cm]{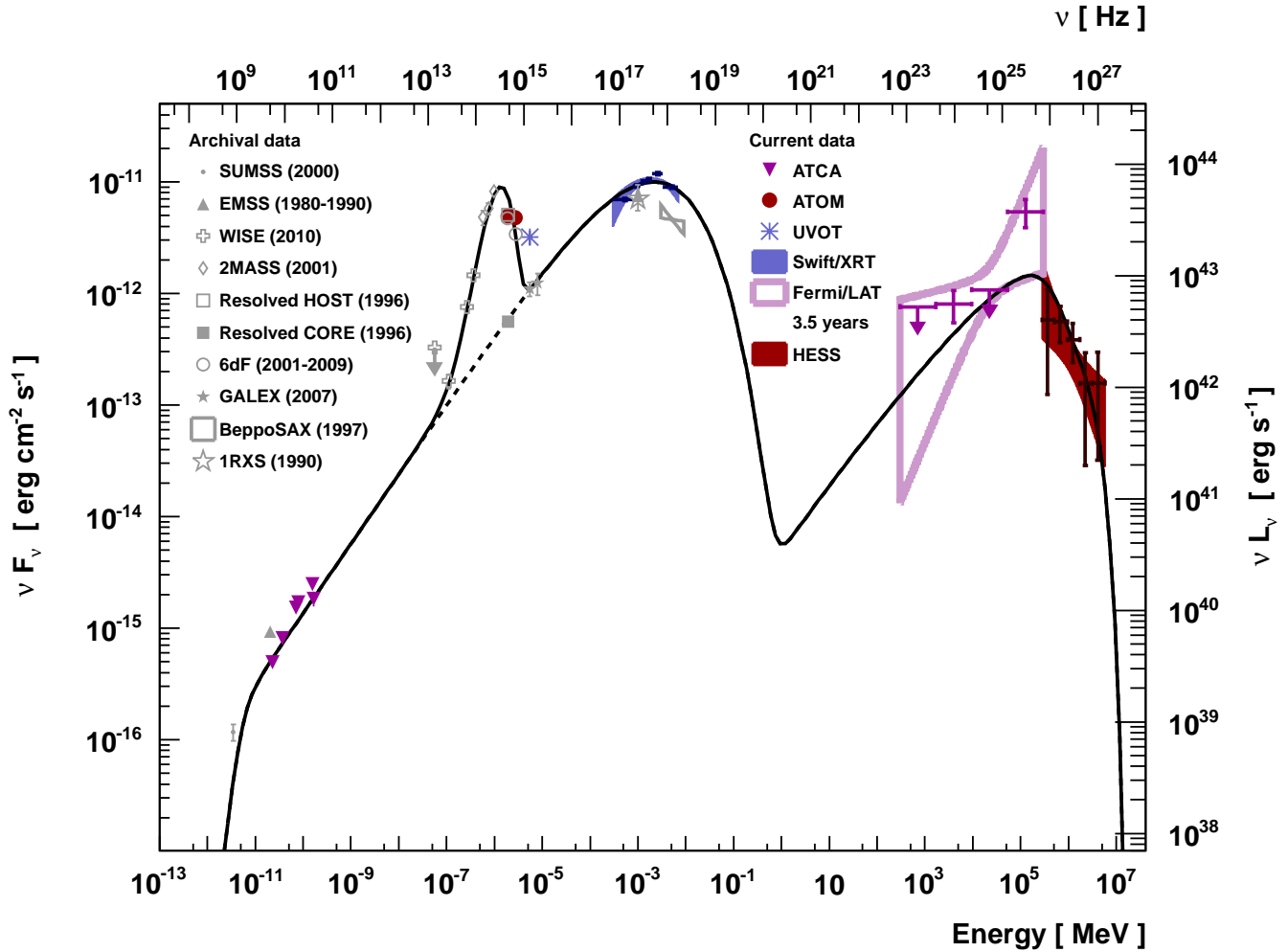}
      \caption{Spectral energy distribution of \OneES. The archival data from radio to X-rays are detailed in the top left legend, the period of observation being indicated between parentheses. 
The observations described in this paper consist of ATOM measurements in the B and R bands (full circles), \swiftuvot\ measurement at 192.8 nm (asterisk symbol) and ATCA measurements at 5.5, 9.0, 17, 19, 38 and 40~GHz (filled downward-pointing triangles).  
\swiftxrt, \hess\ and \fermi\ spectra are represented by the butterflies (filled for the first two, empty for the last one). The data are modeled assuming a black-body emission in the infrared-optical band while the radio and optical emission of the core as well as the X-ray, HE and VHE spectra are described with a one zone homogeneous SSC emission.}
\label{fig:SED}
\end{figure*}

The SED of \OneES\ shown in Fig.~\ref{fig:SED} is derived from a compilation of the data 
analyzed in this paper and of archival data. 
ATCA, ATOM and \swiftuvot\ measurements are described in the top right legend in Fig.~\ref{fig:SED}. 
A conservative approach motivated the use of the \swiftxrt\ spectra corresponding to the minimal and maximal column densities (see Sec.~\ref{swiftXRT}). The uncertainty on the column density is thus practically treated as a systematic effect.

The archival data on \OneES, which are detailed in the top left legend in Fig.~\ref{fig:SED}, are retrieved from on-line databases\footnote{http://vizier.u-strasbg.fr/viz-bin/VizieR, http://tools.asdc.asi.it/SED}. Infrared to UV data, which are dereddened consistently with Sec.~\ref{swiftXRT}, Sec.~\ref{swift} and Sec.~\ref{atom}, are extracted from the Wide-field Infrared survey \citep[WISE,][]{WISE}, from the 2MASS All-Sky Catalog of Point Sources \citep{Cutri}, from the 6dF galaxy survey \citep{2009MNRAS.399..683J} and from the ultraviolet survey performed with \galex\ \citep{2005ApJ...619L...1M}. The UV fluxes measured with \galex\ and with UVOT differ by a factor of three. Both measurements are dereddened and the discrepancy can hardly be explained by the known uncertainties. A high-amplitude variability could explain the difference, though it is barely observed in the other energy bands on the time scale of years. These two points are subsequently not included in the modeling, but are discussed {\it a posteriori} in Sec.~\ref{modeling}. The optical data from the USNO-A2.0 \citep{Monet1998}, the USNO-B1.0 \citep{Monet2003} and the Guide Star \citep[GSC2.3,][]{2006MmSAI..77.1166S} catalogs are not included because of flags indicating a probable association with a nearby star. X-ray and radio archival data are extracted from the Einstein EMSS survey \citep{1990ApJS...72..567G,1991ApJS...76..813S}, the \ROSAT\ All-Sky Bright Source Catalog \citep[1RXS,][]{voges}, the Spectral catlog of \BeppoSAX\ blazars \citep{2005A&A...433.1163D} and the Sydney University Molonglo Sky Survey \citep[SUMSS V2.1,][]{Mauch03}.

\citet{2000A&A...357...91F} imaged the source in the R band and derived an extension of the host galaxy equivalent to $\sim 2~{\rm arsec}$ within the standard $\Lambda$CDM cosmological model ($\Omega_M = 0.27$, $\Omega_\Lambda = 0.73$, $h_0 = 0.71$). The emission of the host galaxy, represented by an open square in Fig.~\ref{fig:SED}, is ten times brighter than the emission of the core represented by the filled square. The fluxes measured by ATOM and from the 6dF galaxy survey are corrected for the limited aperture, yielding 30\% and 20\% greater flux values, respectively. The host galaxy is modeled with a black-body emission, this thermal origin being supported by the absence of polarization and micro-variability \citep{pola}. This emission is constrained by the measurements from WISE, which do not probe synchrotron emission as occurs in brighter TeV BL Lac objects \citep{2011ApJ...740L..48M}. The small amplitude of the variations detected in the ATOM band is fully consistent with the modeling, where the core emission represents {\it a posteriori} 10\% (resp. 20\%) of the total emission in the R (resp. B) band\footnote{Knowing the fraction of the emission coming from the core, a consistent fractional variance of the core flux of $\sim 50\%$ is derived in both R and B ATOM bands.}. Though not strictly contemporaneous, the radio, X-ray HE and VHE data, as well as the emission of the core resolved by \citet{2000A&A...357...91F}, are modeled assuming a non-thermal emission within a standard leptonic synchrotron self Compton scenario.

\subsection{Modeling}\label{modeling}

A canonical, one zone, homogeneous, time independent SSC model is used to interpret the multi wavelength data. A blob of plasma, filled with a constant tangled magnetic field $B$, is modeled by a spherical region of size $R$, and is assumed to move with a bulk Doppler factor $\delta$ \cite[as, e.g., in][]{Band85,Katarz01}. The particle energy distribution (PED) of the electrons is described, as in \citet{Giebels07}, by a power law of index $p$ with an exponential cut off $\displaystyle n(\gamma) = n_0 \gamma^{-p} \mathrm{exp}(-\gamma / \gamma_{\rm cut})$, where $\gamma = E / m_{e}c^2$ is the Lorentz factor of the electrons. The normalization factor of the PED $n_0$ is linked to the kinetic energy density of the electrons in the blob frame according to the equation $ u_{e^-} = m_e c^2 \int_{1}^{+\infty} (\gamma-1)\ n(\gamma) \ \mathrm{d}\gamma$. This variable yields better physical insights into the plasma properties than the normalization factor $n_0$, since it can be directly compared to the magnetic energy density in the blob frame $u_B = B^2 / 8\pi$.

The optical emission is modeled with a black-body spectrum, as in \cite{Katarz03}. The synchrotron self absorption and the internal $\gamma$-$\gamma$ absorption are taken into account according to \citet{Gould} and the approximation of \citet{Coppi}, respectively. The interaction of \g\ rays with the EBL is modeled according to \cite{Fran08}. Finally, the luminosity distance of this source, located at a redshift $z = 0.105$ \citep{Rector00}, is computed within the standard $\Lambda$CDM cosmological model, yielding a luminosity distance $D_L = 479$~Mpc.

Although SSC models are the simplest ones used to explain emission from BL Lac objects, they are usually underconstrained. The SSC model described herein has six parameters, three related to the PED ($n_0$, $\gamma_{\rm cut}$, $p$) and three related to the emission zone ($R$, $B$ and $\delta$). The index of the PED is constrained by the HE index and by the fluxes measured from radio wavelengths to X-rays, including the optical flux of the core measured by \citet{2000A&A...357...91F}. For a fixed index, the amplitude and location of the synchrotron peak are proportional to $n_0 \delta^4 R^3 B^2$ and $B \delta \gamma_{\rm cut}^2$, respectively \citep[e.g.][]{Band85}, and can be fixed using the \swiftxrt\ spectrum. The very low energy part of the spectrum (radio data) is controlled by synchrotron self absorption. This imposes a rather weak constraint since the radio flux, which could arise from larger scale structures, is only an upper limit on the emission of the blob.  Three parameters can hence be tuned to fit the amplitude and location of the inverse Compton peak. Thus, whatever the freedom left by the uncertainty on HE and VHE \g-ray spectra, the model is degenerate. 

To reduce this degeneracy, the inverse Compton bump is reproduced by constraining the magnetic field, the distance to equipartition ($|u_{e^-} / u_B - 1|$) and the Doppler factor. A short distance to equipartition ensures a small energy budget \citep[e.g.,][]{1959ApJ...129..849B}. A small Doppler factor is motivated by the Lorentz factors ($\Gamma\sim 3-4$) inferred from the subluminal motion generally observed in TeV blazars or from the distribution of beamed objects within the BL~Lac / radio galaxy unification scheme \citep[cf. bulk Lorentz factor crisis,][]{HenriSauge}.

In addition to the SSC model parameters, the infrared to optical data impose the amplitude and temperature of the black-body emission, which do not increase the degeneracy of the model. The model of the archival and current data is represented in Fig.~\ref{fig:SED}. The parameters of the SSC model are detailed in Table~\ref{tableFit}. The black-body emission peaks at a temperature $T = 4500 $~K for a total luminosity $\mathcal{L}_{BB} = 3.1\times 10^{44}$~erg~s$^{-1}$. The host galaxy UV emission should be much lower than the flux measured with \swiftuvot, which is hence expected to be of non-thermal origin, but is not matched by our minimal SSC model. This could be taken into account at the expense of adding parameters to the PED or adding extra components \citep[e.g., as in][]{1989MNRAS.237..411S}. 
The bolometric luminosity of the emitting region is $\mathcal{L} = 1.8\times 10^{45}$~erg~s$^{-1}$, ranging from 1\% to 10\% of the Eddington luminosity for fiducial values of the black-hole mass between $10^9$ and $10^8$ solar masses.

\begin{table}
\caption{Parameters of the one zone SSC model shown in Fig.~\ref{fig:SED}. \label{tableFit}}
\centering
\begin{tabular}{ccccccc}
\hline
index $p$ & $\gamma_{\rm min}$&  $\gamma_{\rm cut}$ & $B$ & $\delta$ & $u_{e^-} / u_B$ & $R$    \\
   &  &  & mG &  &  & $\times10^{17}$~cm \\
\hline
1.75 & 1 & $1.0\times10^6$ & 10 & 7 & 45 & 2.4 \\
\hline
\end{tabular}
\end{table}

The parameters of the PED fitting the data are not far from those derived by \cite{Giebels07} for Markarian 421, with electron energies between $\gamma_{\rm min} \sim 1$ and $\gamma_{\rm cut} \sim 10^6$ and a rather hard index $p = 1.75$, smaller than the canonical value of 2 derived in diffusive shock acceleration. The size of the emitting region $R \sim 80$~mpc and the amplitude of the magnetic field $B = 10$~mG are similar to those derived for PKS~2155--304 by \citet{2009ApJ...696L.150A}. A lower limit on the variability time scale on the order of a week can be derived from the size of the emitting region and the Doppler factor $\delta=7$. This value of the Doppler factor sets a limit on the Lorentz factor of the emitting region $\Gamma \geq \delta / 2 = 3.5$. The VHE data cannot be reproduced with a system in equipartition and require a ratio as large as $u_{e^-} / u_B = 45$. Such deviations from equipartition in favor of the particles are not unusual when modeling HBLs \citep[see, e.g.,][for Mrk~421 and Mrk~501]{2012arXiv1205.5237M}. A break in the index of $\Delta p = 1$ is expected at the electron energy for which the cooling time (here given by the synchrotron loss rate) equals the time needed to escape the region, typically $R/c$ to $R/(c/3)$ \citep[cf., e.g.,][]{Tavecchio}. This reads $t_{\rm cool} = \left[ {4 \over 3}\ {{\sigma_T c} \over {m_e c^2}} \ \gamma_{\rm break} \ u_{B} \right]^{-1} \sim R/c$ and for the parameters considered herein the break energy is on the order of the cut off in the PED, thus not affecting the self consistency of the modeling.

\section{Conclusion} \label{ccl}

The \hess\ collaboration reports the discovery of the blazar \OneES\ in the VHE \g-ray domain. 
In spite of being one of the faintest VHE \g-ray sources ever detected, with a differential flux at 1~TeV equivalent to 0.5\% of the Crab nebula differential flux, 
the long observation campaign on its neighbor \cena\ unveiled VHE \g-ray emission from \OneES\ at the $\sim6\sigma$ level. 
The analysis of 3.5 years of data from \fermi\ brought to light a HE spectrum that is one of the hardest derived for a blazar 
by the \fermi\ Collaboration, though with a large uncertainty on the index. The combination of these HE and VHE spectra together with ATCA, ATOM and \swift\ measurements 
allows for the first time the broad band spectral energy distribution of this HBL type blazar to be investigated. 
A black-body emission models the flux of the host galaxy and a simple SSC scenario reproduces the non-thermal emission in the radio, X-ray, HE and VHE bands. 

After 3.5 years of observations with \fermi\ and intensive campaigns with \hess, the extragalactic sky begins to reveal 
sources as faint as few thousandths of the Crab nebula flux. The long-term sky monitoring with \fermi\ combined with the next-generation
 Cherenkov observatory, CTA \citep{2011ExA....32..193A}, will be the important ingredients to reveal a broad picture of blazars' 
HE and VHE behaviour.

\section*{Acknowledgments}
{\footnotesize
The support of the Namibian authorities and of the University of
Namibia in facilitating the construction and operation of H.E.S.S.
is gratefully acknowledged, as is the support by the German
Ministry for Education and Research (BMBF), the Max Planck
Society, the French Ministry for Research, the CNRS-IN2P3 and the
Astroparticle Interdisciplinary Programme of the CNRS, the U.K.
Particle Physics and Astronomy Research Council (PPARC), the IPNP
of the Charles University, the South African Department of Science
and Technology and National Research Foundation, and by the
University of Namibia. We appreciate the excellent work of the
technical support staff in Berlin, Durham, Hamburg, Heidelberg,
Palaiseau, Paris, Saclay, and in Namibia in the construction and
operation of the equipment. 

The \textsl{Fermi} LAT Collaboration acknowledges generous ongoing support
from a number of agencies and institutes that have supported both the
development and the operation of the LAT as well as scientific data analysis.
These include the National Aeronautics and Space Administration and the
Department of Energy in the United States, the Commissariat \`a l'Energie Atomique
and the Centre National de la Recherche Scientifique / Institut National de Physique
Nucl\'eaire et de Physique des Particules in France, the Agenzia Spaziale Italiana
and the Istituto Nazionale di Fisica Nucleare in Italy, the Ministry of Education,
Culture, Sports, Science and Technology (MEXT), High Energy Accelerator Research
Organization (KEK) and Japan Aerospace Exploration Agency (JAXA) in Japan, and
the K.~A.~Wallenberg Foundation, the Swedish Research Council and the
Swedish National Space Board in Sweden.

Additional support for science analysis during the operations phase is gratefully
acknowledged from the Istituto Nazionale di Astrofisica in Italy and the Centre National d'\'Etudes Spatiales in France.

The Australia Telescope Compact Array is part of the Australia
Telescope National Facility which is funded by the Commonwealth of
Australia for operation as a National Facility managed by CSIRO. This
research was funded in part by NASA through \textsl{Fermi} Guest Investigator
grant NNH09ZDA001N (proposal number 31263). This research was
supported by an appointment to the NASA Postdoctoral Program at the
Goddard Space Flight Center, administered by Oak Ridge Associated
Universities through a contract with NASA.

This research has made use of the NASA/IPAC Extragalactic Database (NED) 
which is operated by the Jet Propulsion Laboratory, California Institute of Technology, 
under contract with the National Aeronautics and Space Administration.

This research has made use of the VizieR catalog access tool, CDS, Strasbourg, France.

We are grateful to Kim Page for her help on UVOT analysis issues.
}
\bibliographystyle{mn2e}
\bibliography{1ES1312_HESS_Fermi_MNRAS.bbl}

\begin{thebibliography}{}

\bibitem[\protect\citeauthoryear{{Abdo}, {Ackermann}, {Agudo}, {Ajello},
  {Aller}, {Aller}, {Angelakis}, {Arkharov}, {Axelsson}, {Bach} \& et
  al.}{{Abdo} et~al.}{2010}]{SEDfermiBlazars}
{Abdo} A.~A.,  {Ackermann} M.,  {Agudo} I.,    et al., 2010, \apj, 716, 30

\bibitem[\protect\citeauthoryear{{Abdo}, {Ackermann}, {Ajello}, {Allafort},
  {Antolini}, {Atwood}, {Axelsson}, {Baldini}, {Ballet}, {Barbiellini} \& et
  al.}{{Abdo} et~al.}{2010}]{1FGL}
{Abdo} A.~A.,  {Ackermann} M.,  {Ajello} M.,    et al., 2010, \apjs, 188, 405

\bibitem[\protect\citeauthoryear{{Abdo}, {Ackermann}, {Ajello}, {Allafort},
  {Antolini}, {Atwood}, {Axelsson}, {Baldini} et~al.,}{{Abdo}
  et~al.}{2010}]{2010ApJ...715..429A}
{Abdo} A.~A.,  {Ackermann} M.,  {Ajello} M.,    et~al., 2010, \apj, 715,  429

\bibitem[\protect\citeauthoryear{{Abdo}, {Ackermann}, {Ajello}, {Atwood},
  {Axelsson}, {Baldini}, {Ballet}, {Barbiellini}, {Bastieri}, {Bechtol},
  {Bellazzini}, {Berenji} et~al.,}{{Abdo} et~al.}{2010}]{2010ApJ...710.1271A}
{Abdo} A.~A.,  {Ackermann} M.,  {Ajello} M.,    et~al., 2010, \apj, 710, 1271

\bibitem[\protect\citeauthoryear{{Abdo}, {Ackermann}, {Ajello}, {Atwood},
  {Baldini}, {Ballet}, {Barbiellini}, {Bastieri} et~al.,}{{Abdo}
  et~al.}{2010}]{Cenalobes}
{Abdo} A.~A.,  {Ackermann} M.,  {Ajello} M.,    et~al., 2010, Science,  328, 725

\bibitem[\protect\citeauthoryear{Abdo et~al.,}{Abdo  et~al.}{2009}]{BSL}
Abdo A.~A.,  et~al., 2009, Astrophys.J.Suppl., 183, 46

\bibitem[\protect\citeauthoryear{{Ackermann}, {Ajello}, {Albert}, {Allafort},
  {Atwood}, {Axelsson}, {Baldini}, {Ballet} et~al.,}{{Ackermann}
  et~al.}{2012}]{2012ApJS..203....4A}
{Ackermann} M.,  {Ajello} M.,  {Albert} A.,    et~al., 2012, \apjs, 203, 4

\bibitem[\protect\citeauthoryear{{Ackermann}, {Ajello}, {Allafort}, {Antolini},
  {Atwood}, {Axelsson}, {Baldini}, {Ballet} et~al.,}{{Ackermann}
  et~al.}{2011}]{2011ApJ...743..171A}
{Ackermann} M.,  {Ajello} M.,  {Allafort} A.,    et~al., 2011, \apj, 743, 171

\bibitem[\protect\citeauthoryear{{Actis}, {Agnetta}, {Aharonian},
  {Akhperjanian}, {Aleksi{\'c}}, {Aliu}, {Allan}, {Allekotte}, {Antico},
  {Antonelli} \& et al.}{{Actis} et~al.}{2011}]{2011ExA....32..193A}
{Actis} M.,  {Agnetta} G.,  {Aharonian} F.,    et al., 2011, Experimental Astronomy, 32, 193

\bibitem[\protect\citeauthoryear{{Aharonian}, {Akhperjanian}, {Anton}, {Barres
  de Almeida}, {Bazer-Bachi}, {Becherini}, {Behera}, {Bernl{\"o}hr}, {Boisson},
  {Bochow} \& et al.}{{Aharonian} et~al.}{2009}]{2009ApJ...696L.150A}
{Aharonian} F.,  {Akhperjanian} A.~G.,  {Anton} G.,    et al. (H.E.S.S. Collaboration), 2009, \apjl, 696, L150

\bibitem[\protect\citeauthoryear{{Aharonian}, {Akhperjanian}, {Anton}, {de
  Almeida}, {Bazer-Bachi}, {Becherini} et~al.,}{{Aharonian}
  et~al.}{2009}]{HESS_CenA}
{Aharonian} F.,  {Akhperjanian} A.~G.,  {Anton} G.,    et~al. (H.E.S.S. Collaboration), 2009, \apjl, 695, L40

\bibitem[\protect\citeauthoryear{{Aharonian}, {Akhperjanian}, {Bazer-Bachi},
  {Beilicke}, {Benbow}, {Berge}, {Bernl{\"o}hr} et~al.,}{{Aharonian}
  et~al.}{2006}]{aha2006}
{Aharonian} F.,  {Akhperjanian} A.~G.,  {Bazer-Bachi} A.~R.,    et~al. (H.E.S.S. Collaboration), 2006, \aap, 457, 899

\bibitem[\protect\citeauthoryear{{Aharonian}}{{Aharonian}}{2000}]{2000NewA....%
5..377A}
{Aharonian} F.~A.,  2000, \na, 5, 377

\bibitem[\protect\citeauthoryear{{Andruchow}, {Romero} \&
  {Cellone}}{{Andruchow} et~al.}{2005}]{pola}
{Andruchow} I.,  {Romero} G.~E.,    {Cellone} S.~A.,  2005, \aap, 442, 97

\bibitem[\protect\citeauthoryear{{Atwood}, {Abdo}, {Ackermann}, {Althouse},
  {Anderson}, {Axelsson}, {Baldini}, {Ballet}, {Band}, {Barbiellini} \& et
  al.}{{Atwood} et~al.}{2009}]{atw09}
{Atwood} W.~B.,  {Abdo} A.~A.,  {Ackermann} M.,    et al., 2009, \apj, 697, 1071

\bibitem[\protect\citeauthoryear{{Band} \& {Grindlay}}{{Band} \&
  {Grindlay}}{1985}]{Band85}
{Band} D.~L.,  {Grindlay} J.~E.,  1985, \apj, 298, 128

\bibitem[\protect\citeauthoryear{{Becherini}, {Djannati-Ata{\"i}}, {Marandon},
  {Punch} \& {Pita}}{{Becherini} et~al.}{2011}]{Becherini}
{Becherini} Y.,  {Djannati-Ata{\"i}} A.,  {Marandon} V.,    et al., 2011, Astroparticle Physics, 34, 858

\bibitem[\protect\citeauthoryear{{Berge}, {Funk} \& {Hinton}}{{Berge}
  et~al.}{2007}]{2007A&A...466.1219B}
{Berge} D.,  {Funk} S.,    {Hinton} J.,  2007, \aap, 466, 1219

\bibitem[\protect\citeauthoryear{{Bernl{\"o}hr}, {Carrol}, {Cornils},
  {Elfahem}, {Espigat}, {Gillessen}, {Heinzelmann} et~al.,}{{Bernl{\"o}hr}
  et~al.}{2003}]{Bernlohr}
{Bernl{\"o}hr} K.,  {Carrol} O.,  {Cornils} R.,    et~al., 2003, Astroparticle Physics,  20, 111

\bibitem[\protect\citeauthoryear{{Bessell}}{{Bessell}}{1990}]{Bessel90}
{Bessell} M.~S.,  1990, \pasp, 102, 1181

\bibitem[\protect\citeauthoryear{{Burbidge}}{{Burbidge}}{1959}]{1959ApJ...129.%
.849B}
{Burbidge} G.~R.,  1959, \apj, 129, 849

\bibitem[\protect\citeauthoryear{{Burrows}, {Hill}, {Nousek}, {Kennea},
  {Wells}, {Osborne}, {Abbey}, {Beardmore} et~al.,}{{Burrows}
  et~al.}{2005}]{2005SSRv..120..165B}
{Burrows} D.~N.,  {Hill} J.~E.,  {Nousek} J.~A.,    et~al., 2005, \ssr, 120,  165

\bibitem[\protect\citeauthoryear{{Coppi} \& {Blandford}}{{Coppi} \&
  {Blandford}}{1990}]{Coppi}
{Coppi} P.~S.,  {Blandford} R.~D.,  1990, \mnras, 245, 453

\bibitem[\protect\citeauthoryear{{Cutri}, {Skrutskie}, {van Dyk}, {Beichman},
  {Carpenter}, {Chester} et~al.,}{{Cutri} et~al.}{2003}]{Cutri}
{Cutri} R.~M.,  {Skrutskie} M.~F.,  {van Dyk} S.,    et~al., 2003, {2MASS All Sky Catalog of point sources.}

\bibitem[\protect\citeauthoryear{{de Naurois} \& {Rolland}}{{de Naurois} \&
  {Rolland}}{2009}]{Naurois}
{de Naurois} M.,  {Rolland} L.,  2009, Astroparticle Physics, 32, 231

\bibitem[\protect\citeauthoryear{{Dermer} \& {Schlickeiser}}{{Dermer} \&
  {Schlickeiser}}{1993}]{Dermer1993}
{Dermer} C.~D.,  {Schlickeiser} R.,  1993, \apj, 416, 458

\bibitem[\protect\citeauthoryear{{Donato}, {Sambruna} \& {Gliozzi}}{{Donato}
  et~al.}{2005}]{2005A&A...433.1163D}
{Donato} D.,  {Sambruna} R.~M.,    {Gliozzi} M.,  2005, \aap, 433, 1163

\bibitem[\protect\citeauthoryear{{Draine}}{{Draine}}{2003}]{2003ARA&A..41..241%
D}
{Draine} B.~T.,  2003, \araa, 41, 241

\bibitem[\protect\citeauthoryear{{Falomo} \& {Ulrich}}{{Falomo} \&
  {Ulrich}}{2000}]{2000A&A...357...91F}
{Falomo} R.,  {Ulrich} M.-H.,  2000, \aap, 357, 91

\bibitem[\protect\citeauthoryear{{Feldman} \& {Cousins}}{{Feldman} \&
  {Cousins}}{1998}]{Feldman}
{Feldman} G.~J.,  {Cousins} R.~D.,  1998, \prd, 57, 3873

\bibitem[\protect\citeauthoryear{{Franceschini}, {Rodighiero} \&
  {Vaccari}}{{Franceschini} et~al.}{2008}]{Fran08}
{Franceschini} A.,  {Rodighiero} G.,    {Vaccari} M.,  2008, \aap, 487, 837

\bibitem[\protect\citeauthoryear{{Funk}, {Hermann}, {Hinton}, {Berge},
  {Bernl{\"o}hr}, {Hofmann}, {Nayman}, {Toussenel} \& {Vincent}}{{Funk}
  et~al.}{2004}]{Funk}
{Funk} S.,  {Hermann} G.,  {Hinton} J.,    et al.,  2004,  Astroparticle Physics, 22, 285

\bibitem[\protect\citeauthoryear{{Giebels}, {Dubus} \& {Kh{\'e}lifi}}{{Giebels}
  et~al.}{2007}]{Giebels07}
{Giebels} B.,  {Dubus} G.,    {Kh{\'e}lifi} B.,  2007, \aap, 462, 29

\bibitem[\protect\citeauthoryear{{Gioia}, {Maccacaro}, {Schild}, {Wolter},
  {Stocke}, {Morris} \& {Henry}}{{Gioia} et~al.}{1990}]{1990ApJS...72..567G}
{Gioia} I.~M.,  {Maccacaro} T.,  {Schild} R.~E.,    et al.,  1990, \apjs, 72, 567

\bibitem[\protect\citeauthoryear{{Godet}, {Beardmore}, {Abbey}, {Osborne},
  {Cusumano}, {Pagani}, {Capalbi}, {Perri}, {Page}, {Burrows}, {Campana},
  {Hill}, {Kennea} \& {Moretti}}{{Godet} et~al.}{2009}]{2009A&A...494..775G}
{Godet} O.,  {Beardmore} A.~P.,  {Abbey} A.~F.,    et al.,  2009,  \aap, 494, 775

\bibitem[\protect\citeauthoryear{{Gould}}{{Gould}}{1979}]{Gould}
{Gould} R.~J.,  1979, \aap, 76, 306

\bibitem[\protect\citeauthoryear{{G{\"u}ver} \& {{\"O}zel}}{{G{\"u}ver} \&
  {{\"O}zel}}{2009}]{2009MNRAS.400.2050G}
{G{\"u}ver} T.,  {{\"O}zel} F.,  2009, \mnras, 400, 2050

\bibitem[\protect\citeauthoryear{{Hauser}, {M{\"o}llenhoff}, {P{\"u}hlhofer},
  {Wagner}, {Hagen} \& {Knoll}}{{Hauser} et~al.}{2004}]{Haus04}
{Hauser} M.,  {M{\"o}llenhoff} C.,  {P{\"u}hlhofer} G.,    et al.,  2004, Astronomische Nachrichten, 325, 659

\bibitem[\protect\citeauthoryear{{Henri} \& {Saug{\'e}}}{{Henri} \&
  {Saug{\'e}}}{2006}]{HenriSauge}
{Henri} G.,  {Saug{\'e}} L.,  2006, \apj, 640, 185

\bibitem[\protect\citeauthoryear{{Hinshaw}, {Weiland}, {Hill}, {Odegard},
  {Larson}, {Bennett}, {Dunkley}, {Gold} et~al.,}{{Hinshaw}
  et~al.}{2009}]{hinshaw2009}
 {Hinshaw} G.,  {Weiland} J.~L.,  {Hill} R.~S.,    et~al., 2009, \apjs, 180, 225

\bibitem[\protect\citeauthoryear{{Johannesson}, {Moskalenko}, {Digel} \& {for
  the Fermi LAT Collaboration}}{{Johannesson}
  et~al.}{2010}]{2010arXiv1002.0081J}
{Johannesson} G.,  {Moskalenko} I.,  {Digel} S.,    {for the Fermi LAT
  Collaboration} 2010, arXiv:1002.0081

\bibitem[\protect\citeauthoryear{{Jones}, {Read}, {Saunders}, {Colless},
  {Jarrett}, {Parker}, {Fairall}, {Mauch} et~al.,}{{Jones}
  et~al.}{2009}]{2009MNRAS.399..683J}
{Jones} D.~H.,  {Read} M.~A.,  {Saunders} W.,    et~al., 2009, \mnras, 399,  683

\bibitem[\protect\citeauthoryear{{Kalberla}, {Burton}, {Hartmann}, {Arnal},
  {Bajaja}, {Morras} \& {Poeppel}}{{Kalberla} et~al.}{2005}]{Kalberla2005}
{Kalberla} P.~M.~W.,  {Burton} W.~B.,  {Hartmann} D.,    et al.,  2005, VizieR Online Data Catalog,  8076, 0

\bibitem[\protect\citeauthoryear{{Katarzy{\'n}ski}, {Sol} \&
  {Kus}}{{Katarzy{\'n}ski} et~al.}{2001}]{Katarz01}
{Katarzy{\'n}ski} K.,  {Sol} H.,    {Kus} A.,  2001, \aap, 367, 809

\bibitem[\protect\citeauthoryear{{Katarzy{\'n}ski}, {Sol} \&
  {Kus}}{{Katarzy{\'n}ski} et~al.}{2003}]{Katarz03}
{Katarzy{\'n}ski} K.,  {Sol} H.,    {Kus} A.,  2003, \aap, 410, 101

\bibitem[\protect\citeauthoryear{{Laurent-Muehleisen}, {Kollgaard},
  {Feigelson}, {Brinkmann} \& {Siebert}}{{Laurent-Muehleisen}
  et~al.}{1999}]{1999ApJ...525..127L}
{Laurent-Muehleisen} S.~A.,  {Kollgaard} R.~I.,  {Feigelson} E.~D.,
  et al.,  1999, \apj, 525, 127

\bibitem[\protect\citeauthoryear{{Lemoine-Goumard}, {Degrange} \&
  {Tluczykont}}{{Lemoine-Goumard} et~al.}{2006}]{2006APh....25..195L}
{Lemoine-Goumard} M.,  {Degrange} B.,    {Tluczykont} M.,  2006, Astroparticle
  Physics, 25, 195

\bibitem[\protect\citeauthoryear{{Mankuzhiyil {et al.}}}{{Mankuzhiyil {et
  al.}}}{2012}]{2012arXiv1205.5237M}
{Mankuzhiyil} N., et al., 2012, \apj, 753, 154

\bibitem[\protect\citeauthoryear{{Mannheim}}{{Mannheim}}{1993}]{1993A&A...269.%
..67M}
{Mannheim} K.,  1993, \aap, 269, 67

\bibitem[\protect\citeauthoryear{{Mao}}{{Mao}}{2011}]{2011NewA...16..503M}
{Mao} L.~S.,  2011, \na, 16, 503

\bibitem[\protect\citeauthoryear{{Martin}, {Fanson}, {Schiminovich},
  {Morrissey}, {Friedman}, {Barlow} et~al.,}{{Martin}
  et~al.}{2005}]{2005ApJ...619L...1M}
{Martin} D.~C.,  {Fanson} J.,  {Schiminovich} D.,    et~al., 2005, \apjl, 619, L1

\bibitem[\protect\citeauthoryear{{Massaro}, {D'Abrusco}, {Ajello}, {Grindlay}
  \& {Smith}}{{Massaro} et~al.}{2011}]{2011ApJ...740L..48M}
{Massaro} F.,  {D'Abrusco} R.,  {Ajello} M.,  et al.,  2011, \apjl, 740, L48

\bibitem[\protect\citeauthoryear{{Massaro}, {Paggi}, {Elvis} \&
  {Cavaliere}}{{Massaro} et~al.}{2011}]{2011ApJ...739...73M}
{Massaro} F.,  {Paggi} A.,  {Elvis} M.,    et al.,  2011, \apj, 739, 73

\bibitem[\protect\citeauthoryear{{Massaro}, {Tramacere}, {Cavaliere}, {Perri}
  \& {Giommi}}{{Massaro} et~al.}{2008}]{2008A&A...478..395M}
{Massaro} F.,  {Tramacere} A.,  {Cavaliere} A.,    et al.,
  2008, \aap, 478, 395

\bibitem[\protect\citeauthoryear{{Mattox}, {Bertsch}, {Chiang}, {Dingus},
  {Digel}, {Esposito}, {Fierro} et~al.,}{{Mattox}
  et~al.}{1996}]{1996ApJ...461..396M}
 {Mattox} J.~R.,  {Bertsch} D.~L.,  {Chiang} J.,    et~al., 1996, \apj, 461, 396

\bibitem[\protect\citeauthoryear{{Mauch}, {Murphy}, {Buttery}, {Curran},
  {Hunstead}, {Piestrzynski}, {Robertson} \& {Sadler}}{{Mauch}
  et~al.}{2003}]{Mauch03}
{Mauch} T.,  {Murphy} T.,  {Buttery} H.~J.,  2003, \mnras, 342,
  1117

\bibitem[\protect\citeauthoryear{{Monet}}{{Monet}}{1998}]{Monet1998}
{Monet} D.~G.,  1998, in American Astronomical Society Meeting Abstracts
  Vol.~30 of Bulletin of the American Astronomical Society, {The 526,280,881
  Objects In The USNO-A2.0 Catalog}.
p.~1427

\bibitem[\protect\citeauthoryear{{Monet}, {Levine}, {Canzian}, {Ables}, {Bird},
  {Dahn}, {Guetter}, {Harris} et~al.,}{{Monet} et~al.}{2003}]{Monet2003}
{Monet} D.~G.,  {Levine} S.~E.,  {Canzian} B.,    et~al., 2003, \aj, 125,  984

\bibitem[\protect\citeauthoryear{{Nolan}, {Abdo}, {Ackermann}, {Ajello},
  {Allafort}, {Antolini}, {Atwood}, {Axelsson}, {Baldini}, {Ballet} \& et
  al.}{{Nolan} et~al.}{2012}]{2FGL}
{Nolan} P.~L.,  {Abdo} A.~A.,  {Ackermann} M.,
    et al., 2012, \apjs, 199, 31

\bibitem[\protect\citeauthoryear{{Padovani} \& {Giommi}}{{Padovani} \&
  {Giommi}}{1995}]{1995ApJ...444..567P}
{Padovani} P.,  {Giommi} P.,  1995, \apj, 444, 567

\bibitem[\protect\citeauthoryear{{Perlman}, {Madejski}, {Georganopoulos},
  {Andersson}, {Daugherty}, {Krolik} et~al.,}{{Perlman} et~al.}{2005}]{Perlman}
{Perlman} E.~S.,  {Madejski} G.,  {Georganopoulos} M.,    et~al., 2005, \apj, 625, 727

\bibitem[\protect\citeauthoryear{{Piron}, {Djannati-Atai}, {Punch}, {Tavernet},
  {Barrau}, {Bazer-Bachi}, {Chounet}, {Debiais}, {Degrange}, {Dezalay},
  {Espigat} et~al.,}{{Piron} et~al.}{2001}]{2001A&A...374..895P}
{Piron} F.,  {Djannati-Atai} A.,  {Punch} M.,    et~al., 2001, \aap, 374, 895

\bibitem[\protect\citeauthoryear{{Poole}, {Breeveld}, {Page}, {Landsman},
  {Holland}, {Roming}, {Kuin}, {Brown} et~al.,}{{Poole}
  et~al.}{2008}]{uvotphot}
 {Poole} T.~S.,  {Breeveld} A.~A.,  {Page} M.~J.,    et~al., 2008,
  \mnras, 383, 627

\bibitem[\protect\citeauthoryear{{Rector}, {Stocke}, {Perlman}, {Morris} \&
  {Gioia}}{{Rector} et~al.}{2000}]{Rector00}
{Rector} T.~A.,  {Stocke} J.~T.,  {Perlman} E.~S.,    et al.,  2000, \aj, 120, 1626

\bibitem[\protect\citeauthoryear{{Roming}, {Kennedy}, {Mason}, {Nousek}, {Ahr},
  {Bingham}, {Broos} et~al.,}{{Roming} et~al.}{2005}]{RomingUVOT}
{Roming} P.~W.~A.,  {Kennedy} T.~E.,  {Mason} K.~O.,    et~al., 2005, \ssr, 120, 95

\bibitem[\protect\citeauthoryear{{Sambruna}, {Maraschi} \& {Urry}}{{Sambruna}
  et~al.}{1996}]{1996ApJ...463..444S}
{Sambruna} R.~M.,  {Maraschi} L.,    {Urry} C.~M.,  1996, \apj, 463, 444

\bibitem[\protect\citeauthoryear{{Schlegel}, {Finkbeiner} \&
  {Davis}}{{Schlegel} et~al.}{1998}]{1998ApJ...500..525S}
{Schlegel} D.~J.,  {Finkbeiner} D.~P.,    {Davis} M.,  1998, \apj, 500, 525

\bibitem[\protect\citeauthoryear{{Sol}, {Pelletier} \& {Asseo}}{{Sol}
  et~al.}{1989}]{1989MNRAS.237..411S}
{Sol} H.,  {Pelletier} G.,    {Asseo} E.,  1989, \mnras, 237, 411

\bibitem[\protect\citeauthoryear{{Spagna}, {Lattanzi}, {McLean}, {Bucciarelli},
  {Carollo}, {Drimmel}, {Greene}, {Morbidelli}, {Pannunzio}, {Sarasso}, {Smart}
  \& {Volpicelli}}{{Spagna} et~al.}{2006}]{2006MmSAI..77.1166S}
{Spagna} A.,  {Lattanzi} M.~G.,  {McLean} B.,    et al.,  2006, \memsai, 77, 1166

\bibitem[\protect\citeauthoryear{{Stecker}, {de Jager} \& {Salamon}}{{Stecker}
  et~al.}{1996}]{1996ApJ...473L..75S}
{Stecker} F.~W.,  {de Jager} O.~C.,    {Salamon} M.~H.,  1996, \apjl, 473, L75

\bibitem[\protect\citeauthoryear{{Stevens}, {Edwards}, {Ojha}, {Kadler},
  {Hungwe}, {Dutka}, {Tingay}, {Macquart}, {Moin}, {Lovell} \&
  {Blanchard}}{{Stevens} et~al.}{2012}]{2012arXiv1205.2403S}
{Stevens} J.,  {Edwards} P.~G.,  {Ojha} R.,    et al.,  2012, arXiv:1205.2403, Proceedings of Fermi and Jansky: Our 
Evolving Understanding of AGN, St Michaels, MD, November 10-12, 2011, edited
  by R. Ojha, D. J. Thompson and C. Dermer, eConf C1111101 (2011)

\bibitem[\protect\citeauthoryear{{Stickel}, {Padovani}, {Urry}, {Fried} \&
  {Kuehr}}{{Stickel} et~al.}{1991}]{1991ApJ...374..431S}
{Stickel} M.,  {Padovani} P.,  {Urry} C.~M., et al.,
  1991, \apj, 374, 431

\bibitem[\protect\citeauthoryear{{Stocke}, {Morris}, {Gioia}, {Maccacaro},
  {Schild}, {Wolter}, {Fleming} \& {Henry}}{{Stocke}
  et~al.}{1991}]{1991ApJS...76..813S}
{Stocke} J.~T.,  {Morris} S.~L.,  {Gioia} I.~M.,    et al.,  1991, \apjs, 76, 813

\bibitem[\protect\citeauthoryear{{Tavecchio}, {Maraschi} \&
  {Ghisellini}}{{Tavecchio} et~al.}{1998}]{Tavecchio}
{Tavecchio} F.,  {Maraschi} L.,    {Ghisellini} G.,  1998, \apj, 509, 608

\bibitem[\protect\citeauthoryear{{Tramacere}, {Giommi}, {Massaro}, {Perri},
  {Nesci}, {Colafrancesco}, {Tagliaferri} et~al.,}{{Tramacere}
  et~al.}{2007}]{2007A&A...467..501T}
{Tramacere} A.,  {Giommi} P.,  {Massaro} E.,    et~al., 2007, \aap, 467, 501

\bibitem[\protect\citeauthoryear{{Urry}, {Sambruna}, {Worrall}, {Kollgaard},
  {Feigelson}, {Perlman} \& {Stocke}}{{Urry}
  et~al.}{1996}]{1996ApJ...463..424U}
{Urry} C.~M.,  {Sambruna} R.~M.,  {Worrall} D.~M.,    et al.,  1996, \apj, 463, 424

\bibitem[\protect\citeauthoryear{{Vaughan}, {Edelson}, {Warwick} \&
  {Uttley}}{{Vaughan} et~al.}{2003}]{Vaughan}
{Vaughan} S.,  {Edelson} R.,  {Warwick} R.~S.,    et al.,  2003, \mnras,  345, 1271

\bibitem[\protect\citeauthoryear{{Vincent}, {Denanca}, {Huppert}, {Manigot},
  {de Naurois}, {Nayman}, {Tavernet} et~al.,}{{Vincent} et~al.}{2003}]{Vincent}
{Vincent} P.,  {Denanca} J.-P.,  {Huppert} J.-F.,    et~al., 2003, in International Cosmic
  Ray Conference Vol.~5 of International Cosmic Ray Conference, {Performance of
  the H.E.S.S. Cameras}.
p.~2887

\bibitem[\protect\citeauthoryear{{Voges}, {Aschenbach}, {Boller},
  {Br{\"a}uninger}, {Briel}, {Burkert}, {Dennerl}, {Englhauser}
  et~al.,}{{Voges} et~al.}{1999}]{voges}
{Voges} W.,  {Aschenbach} B.,  {Boller} T.,    et~al., 1999, \aap, 349,
  389

\bibitem[\protect\citeauthoryear{{Wilson}, {Ferris}, {Axtens}, {Brown},
  {Davis}, {Hampson}, {Leach}, {Roberts} et~al.,}{{Wilson}
  et~al.}{2011}]{2011MNRAS.416..832W}
{Wilson} W.~E.,  {Ferris} R.~H.,  {Axtens} P.,    et~al., 2011, \mnras, 416, 832

\bibitem[\protect\citeauthoryear{{Wolter}, {Comastri}, {Ghisellini}, {Giommi},
  {Guainazzi}, {Maccacaro}, {Maraschi}, {Padovani}, {Raiteri}, {Tagliaferri},
  {Urry} \& {Villata}}{{Wolter} et~al.}{1998}]{1998A&A...335..899W}
{Wolter} A.,  {Comastri} A.,  {Ghisellini} G.,    et al.,  1998, \aap, 335, 899

\bibitem[\protect\citeauthoryear{{Wright}, {Eisenhardt}, {Mainzer}, {Ressler},
  {Cutri}, {Jarrett}, {Kirkpatrick}, {Padgett}, {McMillan}, {Skrutskie}
  et~al.,}{{Wright} et~al.}{2010}]{WISE}
{Wright} E.~L.,  {Eisenhardt} P.~R.~M.,  {Mainzer} A.~K.,    et~al., 2010, \aj, 140, 1868

\end{thebibliography}

\appendix

{\normalsize
\section{\\Performance of H.E.S.S. analysis for a source\\ at large offset and in an acceptance gradient\\ \ }\label{Appendix}

In this appendix, we investigate the reliability of the VHE \g-ray spectrum of \OneES\ as derived from observations taken at large offset angle.
H.E.S.S. observations are mostly performed in {\it wobble} mode, i.e. pointed along a circle of radius $\sim$0.5\dgr\ centered on the target. 
This value is an optimum between a decrease of the radial acceptance for an increasing offset and 
an increase of the number of regions used for background estimation. 
The large number of observations all around the source provides a locally flat acceptance field. 
\cena\ being the target of the observations studied in this paper, 
\OneES\ is located in a strong gradient of acceptance as shown in Fig.~\ref{fig:AccOneES}.

  \begin{figure}
	\centering
  \includegraphics[width=7.6cm]{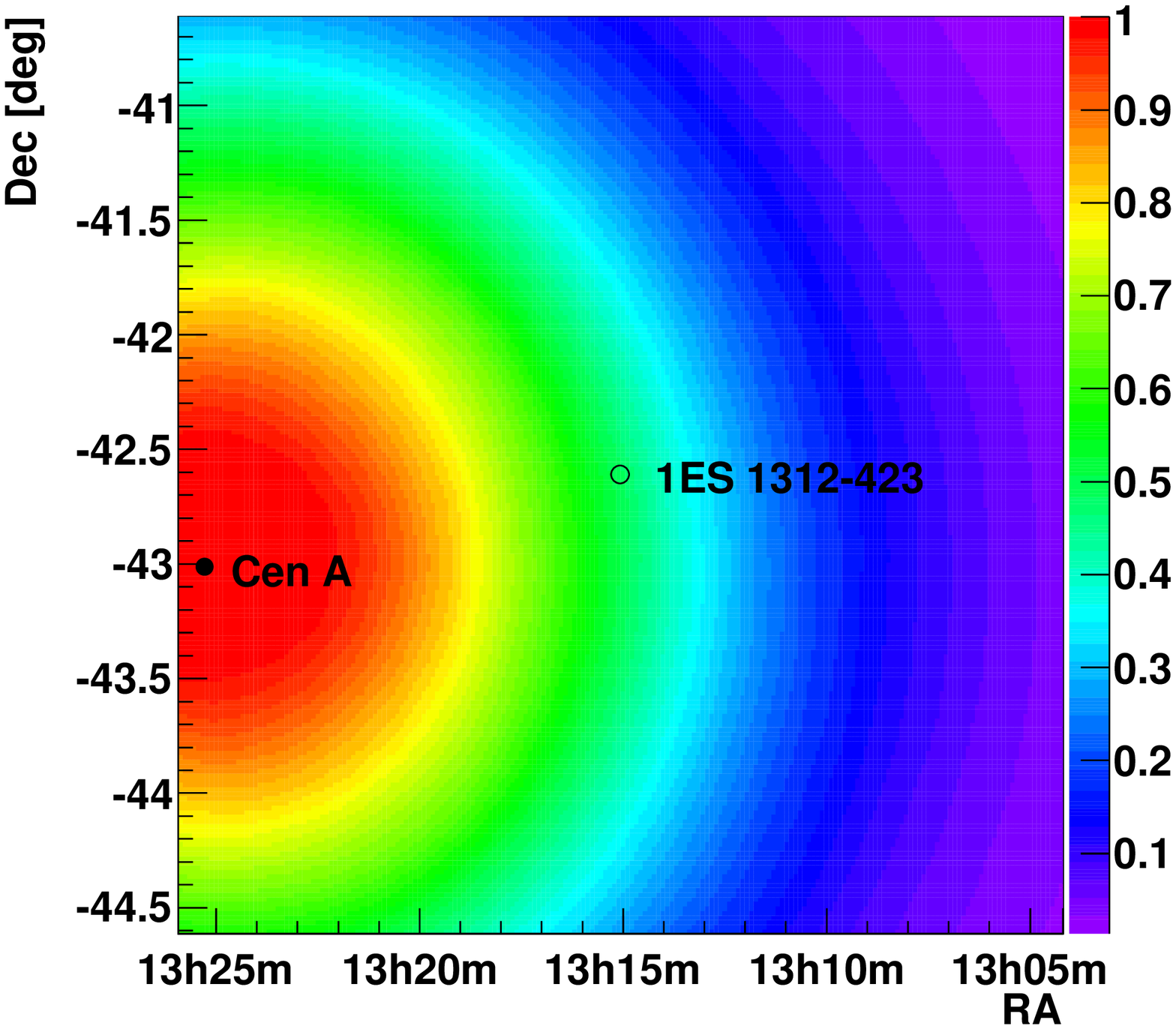}
  \caption{Normalized acceptance map of the \hess\ field of view around \OneES. The map is computed using the measured hadronic events, 
    assumed uniform in the field of view. The maximum value, normalized to 1, corresponds to the location of \cena, 
    target of the runs studied in this paper. \OneES\ is located in a strong gradient of exposure approximately 2\dgr\  
    from the target of observations.}
  \label{fig:AccOneES}
  \end{figure}

\hess\ systematically took data at several offsets from the position of the Crab nebula, the brightest and most studied source in the \hess\ sky \citep{aha2006}, in order to determine the \g-ray acceptance in the field of view. To check the reliability of the analysis at large offsets, 
observations with similarly large offsets from the Crab nebula are selected to reproduce the observational conditions of \OneES. 
The selected dataset corresponds thus to an average offset from the nominal position of $\sim$1.9\dgr\ ($\sim$2\dgr\  for \OneES) 
and a strong gradient of acceptance at the source location, as shown in Fig.~\ref{fig:AccCrab}.

  \begin{figure}
	\centering
  \includegraphics[width=7.6cm]{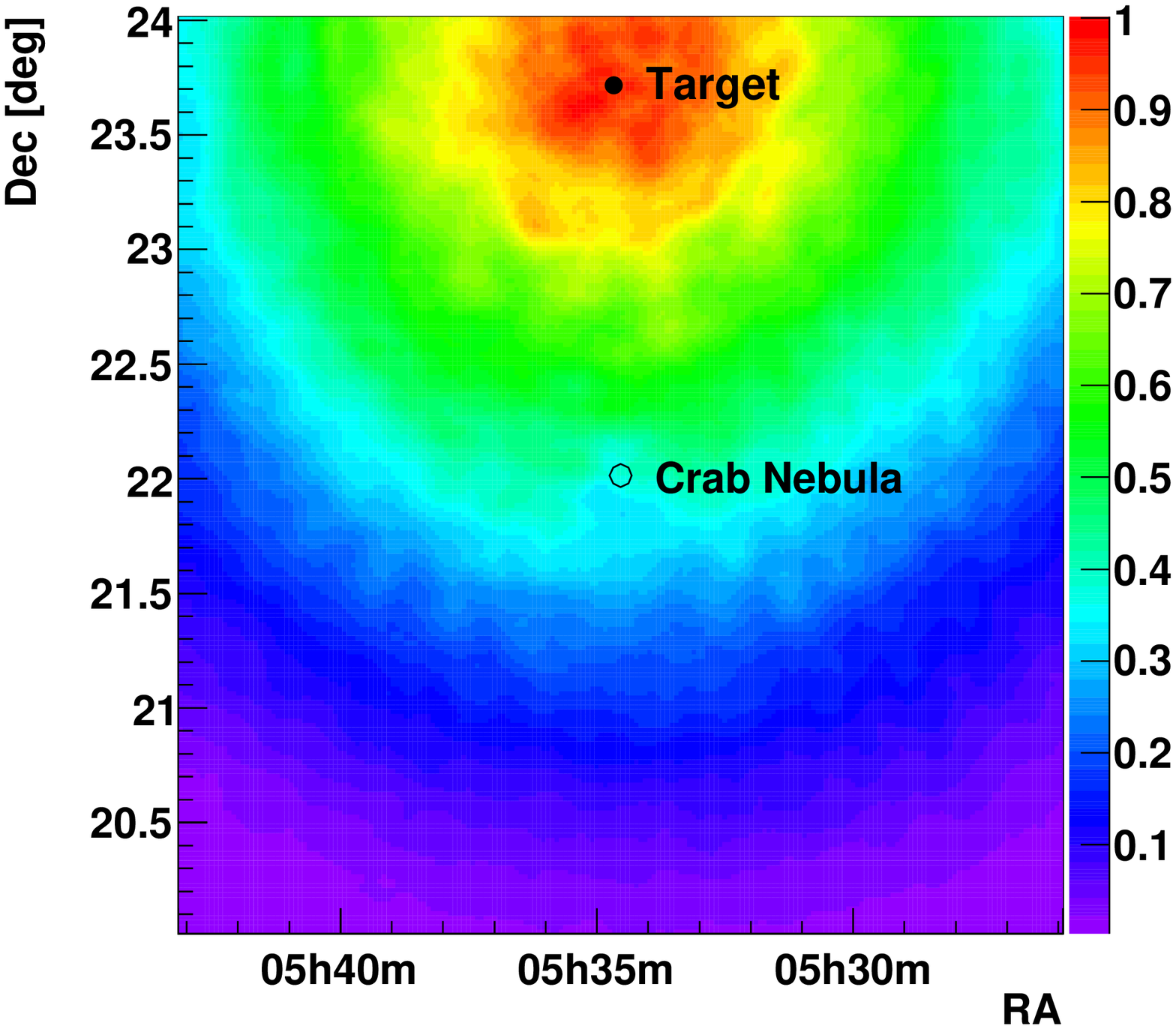}
  \caption{Normalized acceptance map of \hess\ field of view for selected runs on the Crab nebula. 
    The runs are chosen to reproduce the observational conditions of \OneES, 
    locating the Crab nebula in a strong gradient of exposure, approximately 1.9\dgr\  from the hypothetical target.}
  \label{fig:AccCrab}
  \end{figure}

These data are analyzed using the method described in \cite{Becherini}. 
The minimum image intensity of 60 p.e. yields an energy threshold of $680$~GeV for a mean zenith angle of 50$^\circ$. Because of the relatively low statistics on the dataset, a power-law model is preferred to fit the spectrum\footnote{Note that an exponential cut-off power-law model is a better representation for larger datasets, such as studied in \cite{aha2006}.}. 
The parameters of the fit are compared in Fig.~\ref{fig:CrabContours} to those published in \cite{aha2006}, 
where the 1$\sigma$, 2$\sigma$ and 3$\sigma$ confidence contours are plotted in the 
differential-flux-at-1~TeV over power-law-index plane. The power-law spectrum obtained with the selected runs is compatible at the $1\sigma$ level with the published spectrum, confirming the reliability of the spectral analysis of the data acquired by \hess\ on \OneES.

   \begin{figure}
   \includegraphics[width=8.8cm]{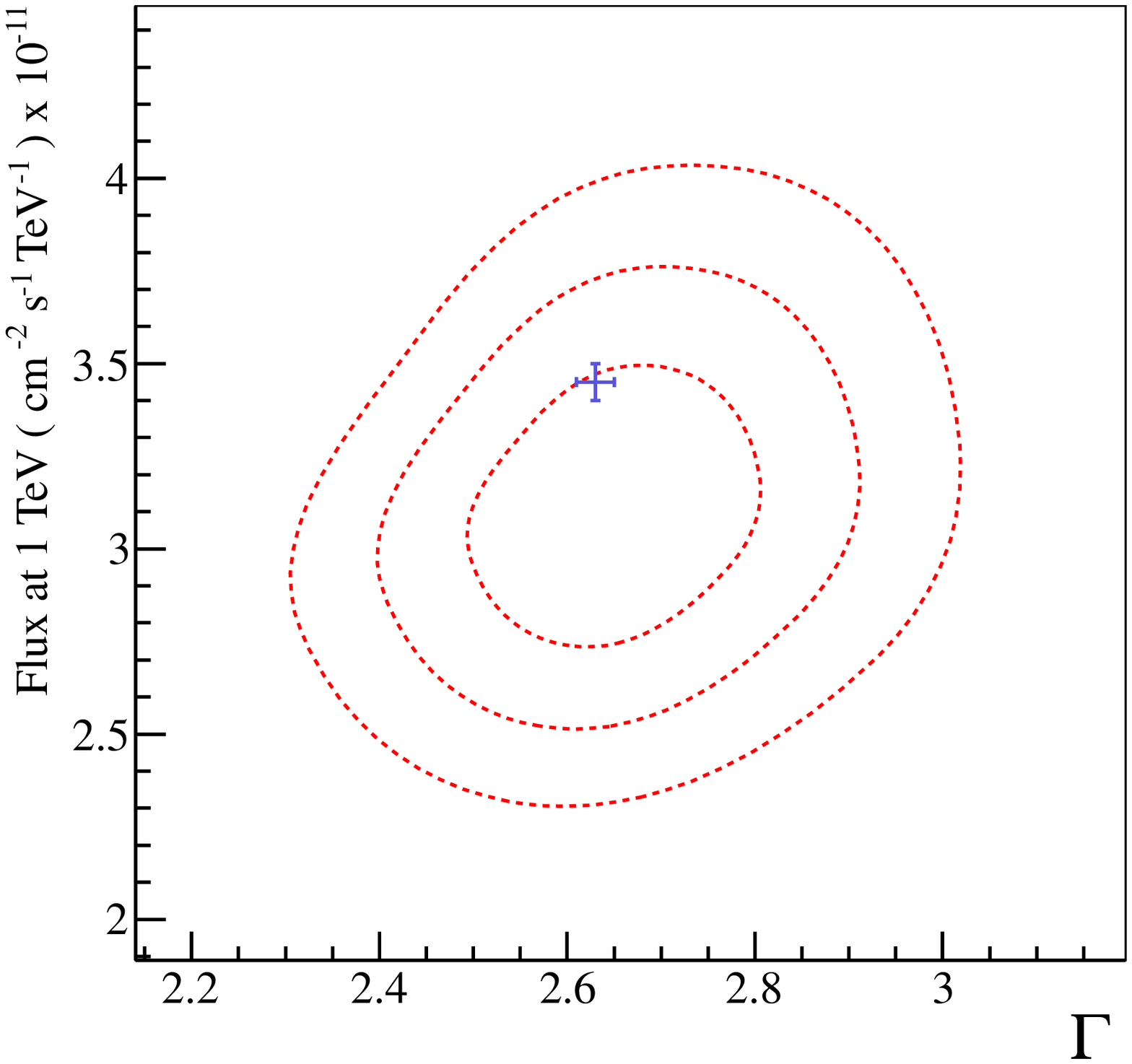}
      \caption{Confidence contours (1$\sigma$ , 2$\sigma$ and 3$\sigma$ levels) of the spectral parameters matching the data from the Crab nebula. 
        The spectrum is fitted with a power law model, characterized by the differential flux at 1~TeV and the power law index $\Gamma$. 
        The best-fit parameters are compatible with the spectrum derived with a larger dataset in \citet{aha2006}, represented by the blue cross.}
   \label{fig:CrabContours}
   \end{figure}  
}

\label{lastpage}

\end{document}